\documentclass{article}
\usepackage{amsfonts,amsbsy,amssymb}
\usepackage{enumerate}
\usepackage{MnSymbol}
\usepackage[T1]{fontenc}
\usepackage[cp1250]{inputenc}

\def\Distr{\hbox{Distr}\,}
\def\spt{\hbox{spt}\,}
\def\sup{\hbox{sup}\,}
\def\tr{\hbox{tr}\,}
\def\Univ{\hbox{Univ}\,}
\def\Kon{\hbox{Kon}\,}
\def\Sym{\hbox{Sym}\,}
\def\diag{\hbox{diag}\,}
\def\into{\hbox{into}\,}
\def\Definition{\vskip 3mm\noindent\textbf{Definition: }}
\def\Theorem{\vskip 3mm\noindent\textbf{Theorem: }}
\def\Remark{\vskip 3mm\noindent\textbf{Remark. }}
\def\Remarks{\vskip 3mm\noindent\textbf{Remarks. }}
\def\Proposition{\vskip 3mm\noindent\textbf{Proposition. }}
\def\Properties{\vskip 3mm\noindent\textbf{Properties. }}
\def\Warning{\vskip 3mm\noindent\textbf{Warning. }}
\def\Proof{\vskip 3mm\noindent\textbf{Proof. }}
\def\Conclusions{\vskip 3mm\noindent\textbf{Conclusions. }}

\begin{document}

\noindent
{\large \textbf{Extended probability theory and quantum mechanics I:\\
non-classical events, partitions, contexts, quadratic probability spaces}}
\vskip 5cm

\textbf{Ji\v r\'{\i}  Sou\v cek}
\vskip 3mm

\textbf{Charles University in Prague}

\textbf{Faculty of Philosophy}
 
\textbf{Ke K\v r\'\i\v zi 8, Prague 5, 158 00}
\vskip 2mm
\textbf{jirka.soucek@gmail.com}
\vfill\eject
{\Large{\textbf{Abstract.}}}
\vskip 3mm

In the paper the basic concepts of extended probability theory are introduced. The basic idea: the concept of an event as a subset of $\Omega$ is replaced with the concept of an event as a partition. The partition is any set of disjoint  non-empty subsets of $\Omega$ (i.e. partition = subset+its decomposition).

Interpretation: elements inside certain part are  in-distinguishable, while elements from different parts are distinguishable. There are incompatible events, e.g $\{\{e_1\},\{e_2\}\}$ and $\{\{e_1,e_2\}\}$. This is logical incompatibility analogical to the impossibility to have and simultaneously not to have the which-way information in the given experiment. The context is the maximal set of mutually compatible events. Each experiment has associated its context. In each context the extended probability is reduced to classical probability. Then the quadratic representation of events, partitions and probability measures is developed. At the end the central concept of quadratic probability spaces (which extend Kolmogorov probability spaces) is defined and studied. In the next  paper it will be shown that quantum mechanics can be represented as the theory of Markov processes in the extended probability theory (Einstein's vision of QM).

\vfill\eject
\section{Introduction}

This paper is the first one from the series of papers concerning  the relation between  Extended Probability Theory (EPT) and Quantum Mechanics (QM).
\vskip 3mm

In this first paper we shall introduce the basis of EPT: non-classical events, incompatibility of events, contexts, extended probability measures, quadratic models of extended events and extended probability measures and at the end the concept of the quadratic probability space.
\vskip 3mm

The basic objects in Probability Theory are events modelled as subsets of $\Omega$, the set of elementary events. Our fundamental idea is to start with the new models for events, where events are partitions in $\Omega$. The partition is the set $A=\{A_\alpha|\alpha\in I\}$ of disjoint non-empty subsets of $\Omega$.
\vskip 3mm

If, $\Omega=\{e_1,e_2,\dots,e_{20}\}$ then partitions are, for example
\begin{enumerate}[(1)]
	\item $\left\{\{e_{16}\},\{e_{1}\},\{e_{7}\},\{e_{4}\},\{e_{20}\}\right\}$
	\item $\left\{\{e_{16},e_{20}\},\{e_{3},e_{5}\}\right\}$
	\item $\left\{\{e_{3},e_{5}\},\{e_7\},\{e_{13},e_{14},e_{15}\},\{e_{17}\}\right\}$
\end{enumerate}

The partition $A$ is classical iff each part $A_\alpha$ is a one-element set: (1) is classical, (2) and (3) are non-classical. The interpretation is the following:
\begin{enumerate}[(i)]
	\item events from the same part are in-distinguishable ($e_{16},e_{20}$ in (2))
	\item events from different parts are distinguishable ($e_{16},e_{3}$ in (2), $e_{16}, e_{20}$ in (1))
\end{enumerate}
\vskip 3mm
Events (1) and (2) cannot be observed in the same experiment, since $e_{16}$ and $e_{20}$ are distinguishable in (1) and in-distinguishable in (2). Such events are called incompatible.

Compatible are (1) and (3), (2) and (3), while (1) and (2) are incompatible.

The context is the maximal set of mutually compatible events. For example, the classical context is the set of all classical events (all classical partitions). In each context there is  the Classical Probability Theory (CPT): i.e. in each context EPT reduces to CPT.
\vskip 3mm

The description of an experiment must contain the definition of the experiment's context (in this way the which-way information enters into physics).
\vskip 3mm

In the paper II we shall show that  EPT contains non-trivial invertible Markov processes and in the paper III we shall introduce the symplectic structure into EPT and then QM can be modeled as the theory of Markov processes in EPT (this will realize the Einstein's vision of QM as a probabilistic theory, like the Brownian motion theory, but in EPT instead of CPT).
\vskip 3mm

Our approach (started in [2], [3]) is principally different from the so-called quantum measure theory (QMT: R.~Sorkin [4], S.~Gudder [5] and others):

\begin{enumerate}[(i)]
	\item The structure of events is completely different in both cases - QMT contain only a part of events contained in EPT
	\item in EPT events have the quadratic structure while in QMT events have linear (=additive) structure
	\item in EPT the probability measure is additive while in QMT is not
	\item in QMT there is no concept of the in-compatibility and no concept of the context: both concepts are necessary for the rational interpretation of QM.
\end{enumerate}
\vskip 3mm

Acknowledgments.

Sincere thanks to my colleagues

M.~Giaquinta (Scuola Normale Superiore, Pisa)

and G.~Modica (University of Florence)

for helping me to understand clearly the central role of partitions.

Many thanks also to J.~Richter and E.~Murtinov\' a (both from Charles University, Prague) for helping me with the preparation of this paper.
\vfill\eject
\section{Classical probability theory and the impossibility to represent QM in it}

The classical probability theory (CPT) contains the following objects and operations:
\vskip 2mm

\begin{enumerate}[(i)]
	\item  $\mathcal E$ is a set of events, it contains the zero event \textbf{0} which never happens (is impossible) and the sure event \textbf{1} which always happens
	\item the operation $\neg :\mathcal E\rightarrow\mathcal E$, the negation, it means the event $\neg A$ happens iff  (=if and only if) $A$ does not happen
	\item the operations $\vee, \wedge:\mathcal E\times\mathcal E\rightarrow \mathcal E$ where $A\vee B$ (=disjunction) happens iff at least one of events $A$, $B$ happens and $A\wedge B$ (=conjunction) happens iff both $A$, $B$ happen
	\item Operations $\vee$, $\wedge$, $\neg$ satisfy the standard commutativity, associativity, distributivity and De Morgan laws
	\item there exists a map $\mathbb F:\mathcal E\rightarrow[0,1]$ where the value $\mathbb F(A)$ denotes the relative frequency of an event of $A$, when this event is (independently) repeated as events $A_1, A_2,\dots $. Let $k_n(A)=$ the number of events from $A_1,\dots,A_n$ which have happened. Then $\mathbb F(A)=\lim \frac{1}{n}k_n(A)$. This means also that the event 
	$$[k_n(A)/n\not{\,\!\!\!\rightarrow}\mathbb F(A)]$$
never happens. We have, of course, $\mathbb F($\textbf{0})=0, $\mathbb F($\textbf{1})=1.
\item We set
$$\mathcal N=\{A\in\mathcal E| \mathbb F(A)=0\}$$
and we can suppose that events $A\in\mathcal N$ never happen.
\end{enumerate}

\vskip 2mm
Usually CPT is considered in the form of the Kolmogorov model. The Kolmogorov model is given as a triple
$$(\Omega,\mathcal A, P)\hbox{ where}$$

\begin{enumerate}[(i)]
	\item $\Omega$ is non-empty set (=the set of elementary events)
	\item $\mathcal A$ (=algebra of events) is a $\sigma$-algebra of subsets of $\Omega$
	\item $P:\mathcal A \rightarrow [0,\infty)$ is the (non-negative) $\sigma$-additive measure on $\Omega$ satisfying $P(\Omega)>0$.
\end{enumerate}
\vskip 2mm

The model for CPT is then defined by the following specifications
\begin{enumerate}[(i)]
	\item $\mathcal E:=\mathcal A$, $\textbf{0}:=\emptyset$, \textbf{1}:=$\Omega$
	\item $A\wedge B:=A\cap B$, $A\vee B:=A\cup B$, $\neg A:=\Omega\backslash A$, $A,B \in\mathcal E$
	\item $\mathbb F(A):=P(A)/P(\Omega)$.
\end{enumerate}

\vskip 2mm
(Of course, usually it is supposed that $P(\Omega)=1$ and then $\mathbb F=P$. But we prefer our formulation where $\mathbb F$ and $P$ are different objects.)

\vskip 2mm
The basic theorem of CPT (the strong Law of Large Numbers) says that the event
$$Z=[k_m(A)/m \not{\,\rightarrow} P(A)/P(\Omega)]$$
has the zero probability, $P(Z)=0$ and thus $Z$ never happens. This shows that the Kolmogorov model for CPT is correct.
\vskip 2mm

In this paper we shall often consider (to simplify the situation) the finite probability spaces, where
$$|\Omega|=\hbox{ the number of elements of } \Omega$$
is finite, say
$$\Omega=\{e_1, e_2,\dots,e_n\}.$$
Clearly, then the relative frequency $\lim k_m(A)/m$ is defined only approximately.
\vskip 2mm

In the case of $\Omega$ finite, there exists a canonical algebra containing all subsets of $\Omega$ 
$$\mathcal A=\mathcal A_\Omega=2^\Omega=\{A|A\subset\Omega\}.$$

In this case the probability measure $P:\mathcal A\rightarrow [0,\infty)$ can be simply identified with the probability distribution
$${\bf p}=(p_1,\dots,p_n),\ p_i=\mathbb F(e_i),\ i=1,\dots, n$$
so that ${\bf p}\in\Distr_n$ where
$$\Distr_n:=\{(q_1,\dots,\ q_n)\in\mathbb R^n | q_1,\dots,q_n\geq0,\ q_1+\dots+q_n=1\}.$$
Then $\mathbb F$ is given  by
$$\mathbb F(A)=\sum\{p_i|e_i\in A\},\ A\subset\Omega.$$
\vskip 5mm

\Definition
The probability transformation $\Phi$ is the map
$$\Phi:\Distr_n\rightarrow\Distr_n$$
which conserves the convex structure of $\Distr_n$, i.e.

$$\Phi(\sum_{i=1}^k\lambda_i{\bf p}^{(i)})=\sum_{i=1}^k\lambda_i\Phi({\bf p}^{(i)})$$
for each ${\bf p}^{(1)},\dots, {\bf p}^{(k)}\in\Distr_{n}, \lambda_1,\dots,\lambda_k\geq0,\ \lambda_1+\dots+\lambda_k=1.$
\vskip 2mm

It is well known that each probability transformation $\Phi$ can be represented as a stochastic matrix $\Phi_{ij}$ such that
$$\Phi(p_1,\dots,p_n)=(\sum\Phi_{1j}p_j,\dots,\sum\Phi_{nj}p_j)$$
$$\Phi_{ij}\geq 0,\ \forall i,j\ \Phi_{1j}+\dots+\Phi_{nj}=1,\ \forall j.$$
\vskip 2mm

Now we can introduce the concept of the non-dissipativity.
\vskip 3mm

\Definition
\begin{enumerate}[(i)]
	\item The probability transformation $\Phi$ is invertible iff the inverse map $\Phi^{-1}:\Distr_n\rightarrow\Distr_n$ exists and $\Phi^{-1}$ is a probability transformation
	\item the probability distribution $(p_1,\dots,\ p_n)\in\Distr_n$ is deterministic iff there exists $i_0$ such that
	$$p_{i_0}=1,\ p_i=0,\ \forall i\ne i_0$$
	\item the probability distribution ${\bf p}$ is non-dissipative iff there exists an invertible probability transformation $\Phi$ such that $\Phi({\bf p})$ is deterministic
	\item $\Phi$ is deterministic iff $\Phi({\bf p})$ is deterministic for each ${\bf p}$ deterministic
		\item $\Phi$ is a permutation iff there exists a permutation $\pi:\{1,\dots,n\}\rightarrow\{1,\dots,n\}$ such that
	$$\Phi(p_1,\dots,p_n)=(p_{\pi(1)},\dots,p_{\pi(n)}).$$
\end{enumerate}
Then we have the following proposition.

\textbf{Proposition}
\begin{enumerate}[(i)]
	\item ${\bf p}\in\Distr_n$ is non-dissipative iff ${\bf p}$ is deterministic
	\item the following properties of $\Phi$ are equivalent
	\begin{enumerate}[(a)]
		\item $\Phi$ is deterministic and one-to-one
		\item $\Phi$ is invertible
		\item $\Phi$ is a permutation
	\end{enumerate}
\end{enumerate}

\Proof Let $\Phi$ be invertible. We shall show that $\bf p$ non-deterministic $\implies$  $\Phi ({\bf p})$  non-deterministic. If $\bf p$ is non-deterministic then there exist ${\bf p_1},{\bf p_2}\in\Distr_n$, ${\bf p_1} \neq {\bf p_2}$ , $0<\lambda<1$ such that ${\bf p}=\lambda {\bf p_1}+(1-\lambda){\bf p_2}$. Then $\Phi({\bf p})=\lambda \Phi({\bf p_1})+(1-\lambda)\Phi({\bf p_2})$ and $\Phi({\bf p_1})\neq \Phi({\bf p_2})$ and this shows that $\Phi({\bf p})$ is non-deterministic.

Thus $\bf p$ deterministic $\implies$  $\Phi({\bf p})$  deterministic.
 
(i) Let $\bf p$ is non-dissipative. Then there exists $\Phi$ invertible such that $\Phi(\bf p)$ is deterministic. Then $\Phi^{-1}(\Phi ({\bf p}))={\bf p}$ is deterministic. 

(ii) (b) $\implies$ (a) $\implies$ (c) $\implies$ (b) 

\vskip 3mm

\Remark It is clear that each constant map $\Phi:\Distr_n\rightarrow\Distr_n$ is the probability transformation. Thus the condition of the invertibility of $\Phi$ in the definition of the non-disipativity of ${\bf p}$ is necessary - otherwise each $\bf p$ would be non-dissipative.
\vskip 5mm

The discrete Markov process (a Markov chain) is the semigroup of probability transformations parametrized by positive integers. It is a set of probability transformations
$$\{\Phi_{s,t}|s,t\in\mathbb N, s>t\}$$
satisfying the chain rule
$$\Phi_{s,t}=\Phi_{s,r}\circ\Phi_{r,t},\ \forall s>r>t,\ s,r,t\in\mathbb N.$$

The Markov process is deterministic iff each probability transformation $\Phi_{s,t}$ is deterministic. This means that if the initial probability  distribution  ${\bf p}(0)$ is deterministic, then each later probability distribution
$${\bf p}(s)=\Phi_{s,0}({\bf p}(0))$$
will be deterministic, too. So that there will be no randomness in this process.
\vskip 2mm
The processes in Quantum Mechanics (QM) have two important properties

\begin{enumerate}[(i)]
	\item they are non-deterministic:\\
	the QM evolution is fundamentally probabilistic, in fact, only probabilities for the future can be predicted. Starting from the deterministic state, the system evolves into non-deterministic states. Only probabilities of results of repeated experiments can be predicted
	\item the evolution in QM is invertible.
\end{enumerate}
\vskip 2mm

These two properties clearly imply that the QM evolution cannot be described as a Markov process in CPT. In fact, the invertibility in CPT implies that the process must be deterministic.
\vskip 2mm

\textbf{Conclusion:} QM cannot be represented as a Markov process in CPT.
\vfill\eject

\section{Non-classical events, irreducibility and compatibility in Extended Probability Theory (EPT).}

In EPT there are two possibilities how to construct new events from elementary (or previously constructed) events:

\begin{enumerate}[(i)]
	\item if we have a subset
	$$A=\{e_{i_1},\dots,e_{i_k}\}\subset\Omega$$
	then the irreducible (or in-distinguishable) union
	$$\sqcup A:=e_{i_1}\sqcup\dots\sqcup e_{i_k}$$
	can be constructed.\\
	\vskip 2mm
	Such events are called irreducible or atomic events (simply atoms).\\
	For $k=1$, $A=\{e_{i_1}\}$ the following notation will be used
	$$\sqcup A=\sqcup\{e_{i_1}\}=e_{i_1}$$
	\vskip 2mm
	Events $e_1=\sqcup\{e_{i_1}\},\dots,e_n=\sqcup\{e_{n}\}$ are called the classical atoms.\\
	\vskip 2mm
	The support of $\sqcup A$ is defined as
	$$\spt(\sqcup A)=A=\{e_{i_1},\dots,e_{i_k}\}\subset\Omega.$$
	\item if we have atoms $a_1=\sqcup A_1,\dots,$ $a_s=\sqcup A_s$ with disjoint supports $\spt a_1=A_1,\dots, \spt a_s=A_s$ then the reducible (or distinguishable) union
	$$a_1\vee a_2\vee\dots\vee a_s$$
can be created.

\vskip 2mm
Thus the process of the formation of events in EPT is two-step: at the first step atoms are formed as irreducible  unions of elementary events and at the second step the reducible unions of disjoint atoms are formed.

(In CPT the process of the formation of events contain only one step: the reducible unions of elementary events are created.)
\vskip 2mm

There are important points which have to be mentioned.

\begin{enumerate}[(i)]
	\item reducible unions are formed only from disjoint atoms. For example forming the reducible union
	$$(e_1\sqcup e_2)\vee(e_2\sqcup e_3)$$
	from atoms $e_1\sqcup e_2$, $e_2\sqcup e_3$ means that $e_2\sqcup e_3$ can "distruct" the irreducibility (in-distinguishability) of the atom $e_1\sqcup e_2$
	\item the formation of irreducible union of non-atomic events leads to a contradiction. For example the "possible" event
	$$e=(e_1\vee e_2)\sqcup e_3$$
	is contradictory, since the reducibility of $e_1\vee e_2$ is in contradiction with the irreducibility of $e$. The events $e_1, e_2, e_1\vee e_2$ are distinguishable from $e_3$, and the irreducibility of $(e_1\vee e_2)\sqcup e_3$ is destroyed.
\end{enumerate}

\end{enumerate}

\vskip 2mm
The classical events are reducible unions of elementary events, or equivalently, the reducible unions of classical atoms. On the other extreme there are non-classical atoms, which are irreducible unions of elementary events.
\vskip 2mm

\Definition
\begin{enumerate}[(i)]
	\item we say that non-empty sets $A_n,\dots,A_s\subset\Omega$ are ortogonal
	$$\bot(A_1,\dots,A_s)$$
	if sets $A_1,\dots,A_s$ are pair-wise disjoint
	\item the set of events in EPT is
	$$\mathcal E_\Omega:=\{(\sqcup A_1)\vee\dots\vee(\sqcup A_s)|A_1,\dots,A_s\subset\Omega,\bot(A_1,\dots,A_s)\}$$
	\item the set of classical events in EPT is
	$$\mathcal E_\Omega^{cl}=\{\vee A|A\subset\Omega\}$$
	more precisely if $A=\{e_{i_1},\dots,e_{i_k}\}$ then 
	$$\vee A=(\sqcup\{e_{i_1}\})\vee\dots\vee(\sqcup\{e_{i_k}\}).$$
	\item the event $E\in\mathcal E_\Omega$ is an irreducible (or atomic) iff there exists $A\subset \Omega$ such that
	$$E=\sqcup A$$
	the set of all irreducible events is denoted by
	$$\mathcal E_\Omega^{irr}=\{\sqcup A | A\subset \Omega\}$$
	\item for each event
	$$e=(\sqcup A_1)\vee\dots\vee(\sqcup A_s)\in\Omega$$
	we set
	$$\spt e=A_1\cup\dots\cup A_s\subset \Omega.$$
\end{enumerate}

\vskip 2mm
Now it is clear what we mean by the term "extended". This means that we introduced into the probability theory a new type of events $(\in\mathcal E_\Omega\backslash\mathcal E_\Omega^{cl})$ which do not exists in CPT.

The classical events form the subset of all events in EPT. The set classical events $\mathcal E_\Omega^{cl}$ is isomorphic to the set of events $\mathcal A_\Omega$ in CPT by
$$(\sqcup\{e_{i_1}\})\vee\dots\vee(\sqcup\{e_{i_k}\})\leftrightarrow e_{i_1}\vee\dots\vee e_{i_k}.$$

The change from the set $\mathcal E_\Omega^{cl}$ to $\mathcal E_\Omega$ of course implies many changes in probability theory. In this and in following papers we shall study consequences of this change.
\vskip 2mm

Now having the extended set of events $\mathcal E_\Omega$ we simply see that not any two events can be observable in a given experiment. For example $e_1\sqcup e_2$ and $e_1\vee e_2$ cannot be both observed in the same experiment: observing $e_1\vee e_2$ we cannot simultaneously observe $e_1\sqcup e_2$, since the reducibility of $e_1\vee e_2$ would contradict to irreducibility of $e_1\sqcup e_2$.

We cannot reduce $e_1\sqcup e_2$ into $e_1$ and $e_2$. This is equivalent to the impossibility simultaneously to have and not to have the which-way information in QM.

This is the purely logical incompatibility.
\vskip 2mm

By the compatibility of two events $e,f\in\mathcal E_\Omega$ we mean that it is possible to observe $e$ and $f$ in the same experiment.
\vskip 2mm

Other examples of incompatible events are:
\begin{gather*}
e_1\sqcup e_2,\ e_1\\
e_1\sqcup e_2,\ e_2\\
e_1\sqcup e_2,\ e_2\sqcup e_3\\
e_1\sqcup e_2,\ e_2\vee e_3\hbox{ etc.}
\end{gather*}

These examples support the following definition

\Definition\begin{enumerate}[(i)]
	\item let $a=\sqcup A,\ b=\sqcup B\in\mathcal E_\Omega$ be two atoms. Atoms $a$ and $b$ are compatible
	$$a\pitchfork b$$
	iff either $a=b$ or $a\bot b$ (i.e. $\spt a\cap\spt b=\emptyset$)
	\item let $e=a_1\vee\dots\vee a_s$, $f=b_1\vee\dots\vee b_r\in\mathcal E_\Omega$, $a_1,\dots,a_s,b_1,\dots,b_r$ are atoms, $\bot(a_1,\dots,a_s)$, $\bot(b_1,\dots,b_r)$\\
	then
	$$e\pitchfork f\hbox{ iff }a_i\pitchfork b_j,\ \forall i=1,\dots,s,\ \forall j=1,\dots,r.$$
\end{enumerate}
\vskip 3mm
This means that two events are compatible if all atoms inside of them are either equal or disjoint.
\vskip 2mm

The inclusion of events is defined only for compatible events
\vfill\eject
\Definition

\noindent
Let $e=a_1\vee\dots\vee a_s,\  f=b_1\vee\dots\vee b_r\in\mathcal E_\Omega$, $a_1,\dots,a_s,b_1,\dots,b_r$ atoms, $\bot (a_1,\dots,a_s),\ \bot(b_1,\dots,b_r)$.\\
Then we set
$$e\leq f$$
iff $\forall i=1,\dots,s$ there exists $j\in\{1,\dots,r\}$ such that $a_i=b_j$ (clearly $e\leq f$ iff $e\pitchfork f$ and $\spt e\subset\spt f$.)
\vskip 2mm

It is possible also to define the irreducible union of two atoms.

\Definition
\begin{enumerate}[(i)]
	\item let $a=\sqcup A$, $b=\sqcup B$ are two atoms from $\mathcal E_\Omega$. Then we set
	$$a\sqcup b:=\sqcup (A\cup B)=\sqcup(\spt a\cup \spt b)$$
	This irreducible union of two atoms creates a new atom.
	\item for each event $e\in\mathcal E_\Omega$ we can define its "irreducible closure" $\sqcup e$ by
	$$\sqcup e:=\sqcup(\spt e).$$
\end{enumerate}

\Remark It is clear that the set of all atoms together with operations $\sqcup,\wedge,\neg$ and elements $\emptyset, \Omega$ form the Boolean algebra if
$$a\wedge b:=\sqcup(\spt a\cap\spt b),$$
$$a\sqcup b:=\sqcup(\spt a\cup\spt b)\hbox{ - as defined above}$$
$$\neg a:=\sqcup(\Omega\backslash\spt a).$$

\vfill\eject
\section{Partitions and events in EPT}

We have seen that the general event $e\in\mathcal E_\Omega$ can be expressed as
$$e=(\sqcup A_1)\vee\dots\vee(\sqcup A_s)$$
where $A_1,\dots,A_s$ are not-empty disjoint subsets of $\Omega$. Classical events are described as subsets of $\Omega$ i.e.
$$e=e_{i_1}\vee\dots\vee e_{i_k}=\vee A,\ A=\{e_{i_1},\dots,e_{i_k}\}\subset\Omega.$$
Thus the main generalization presented here is the change
$$\{\hbox{subsets}\}\rightarrow\{\hbox{partitions}\}$$
where partitions $\{A_1,\dots,A_s\}$ will be defined below.
\vskip 2mm

We shall see that events in EPT are naturally parametrized by partitions and that operations defined on partitions are the key concepts in EPT.
\vskip 2mm

\noindent
\textbf{Warning:} The partition always mean here the (generally) incomplete partition, i.e. in general we have $\cup A_\alpha\ne\Omega$.

\Remark Partitions are naturally considered in the general setting, where $\Omega$ can be any not-empty set,  possibly of any cardinality.

\Definition Let $\Omega$ be any not-emtpy set. A system
$$A=\{A_\alpha|\alpha\in I\}$$
where $I$ is any index set is a partition in $\Omega$ iff
\begin{enumerate}[(i)]
	\item each $A_\alpha$ is a not-empty part of $\Omega,\alpha\in I$
	\item $A_\alpha\cap A_\beta=\emptyset,\ \forall\alpha\ne\beta,\ \alpha,\beta\in I$\\
	i.e. parts $A_\alpha$ are disjoint\\
	(it may happen that $\bigcup_\alpha A_\alpha\ne\Omega$, so that $A$ is an incomplete partition.)\\
\end{enumerate}
The set of all partitions in $\Omega$ will be denoted by $\Pi_\Omega$.

\Definition Let $A=\{A_\alpha|\alpha\in I\}\in\Pi_\Omega$ be a partition in $\Omega$.
\begin{enumerate}[(i)]
	\item $A$ is a classical partition iff\\
	$|A_\alpha|=$the number of elements in $A_\alpha=1,\ \forall\alpha\in I$.\\
	i.e. classical partition is for example
	$$A=\{\{e_{i_1}\},\dots,\{e_{i_k}\}\}.$$
	the set of classical partitions will be denoted $\Pi_\Omega^{cl}$.
	\item $A$ is an irreducible or atomic partition iff
	$$|I|=1\hbox{ i.e., } A=\{A_1\},\ A_1\subset\Omega.$$
	the irreducible partition is, for example\\
	$$A=\{\{e_{i_1},\dots,e_{i_k}\}\}$$
	The set of all irreducible partitions will be denoted $\Pi_\Omega^{irr}$
	\item the support of $A$ is defined by
	$$\spt A=\bigcup_\alpha A_\alpha\subset\Omega.$$
	\item the partition $A$ is complete iff
	$$\spt A=\bigcup_\alpha A_\alpha=\Omega$$
	\item for each partition $A=\{A_\alpha|\alpha\in I\}\in\Pi_\Omega$ we define an irreducible partition $\neg A$ by
	$$\neg A:=\{\Omega\backslash\spt A\}.$$
	\item for $A\in\Pi_\Omega$ we define its irreducible closure by
	$$\sqcup A=\neg\neg A=\{\spt A\}$$
	
\end{enumerate}

\Remark It is clear (and very important) that the concept of a partition is a union of two basic concepts: the concept of a subset and the concept of  a decomposition. The partition can be seen as a decomposition of a subset. This gives the inner structure to subsets (distinguishability or reducibility among elements of it).
\vskip 3mm

In $\Pi_\Omega$ there are natural operations $\wedge$ and $\vee$.

\Definition let $A=\{A_\alpha|\alpha\in I\},\ B=\{B_\beta|\beta\in J\}\in\Pi_\Omega$. Then
\begin{enumerate}[(i)]
	\item we set
	$$A\wedge B:=\{A_\alpha\cap B_\beta|(\alpha,\beta)\in I'\},\hbox{ where } I'=\{(\alpha,\beta)\in I\times J|A_\alpha\cap B_\beta\ne\emptyset\}.$$
	\item if $\spt A\cap\spt B=\emptyset$ then we set
	$$A\vee B:=A\cup B=\{A_\alpha|\alpha\in I\}\cup\{B_\beta|\beta\in J\}.$$
	\item if $\spt A\cap\spt B\ne\emptyset$ then we set
	$$A\vee B:=(A\wedge \neg B)\vee(\neg A\wedge B)\vee(A\wedge B)$$
	using the definition (ii), since supports of $A\wedge\neg B$, $\neg A\wedge B$, $A\wedge B$ are disjoint.
	\item Also the zero partition
	$$\emptyset=\{A_\alpha|\alpha\in I\}, I=\emptyset$$
	is allowed in $\Pi_\Omega$
	\item Two partitions $A,\ B\in\Pi_\Omega$ are orthogonal, $A\bot B$ iff $\spt A$ and $\spt B$ are disjoint.
\end{enumerate}

\Proposition
\begin{enumerate}[(i)]
	\item operations $\wedge$ and $\vee$ in $\Pi_\Omega$ are commutative and associative
	\item the distribution law
	$$A\wedge(B\vee C)=(A\wedge B)\vee(A\wedge C)$$
	holds in $\Pi_\Omega$.
	\item the distribution law
	$$A\vee(B\wedge C)=(A\vee B)\wedge(A\vee C)$$
	does not hold in $\Pi_\Omega$.
\end{enumerate}
\Proof The proof is not difficult and will be given elsewhere.
\vskip 3mm

The compatibility (and incompatibility) of partitions will be the central concept in the sequel.

\Definition Let $A=\{A_\alpha|\alpha\in I\},\  B=\{B_\beta|\beta\in J\}\in\Pi_\Omega$.
\begin{enumerate}[(i)]
	\item $A$ and $B$ are compatible, $A\pitchfork B$ iff $\forall\alpha\in I\ \forall\beta\in J$ we have
	$$\hbox{either }A_\alpha=B_\beta\hbox{ or }A_\alpha\cap B_\beta=\emptyset$$
	\item We set $A\leq B$ iff $A\subset B$ as sets, i.e. $\forall\alpha\exists\beta$ such that $A_\alpha=B_\beta$. (Clearly $A\pitchfork B\Leftrightarrow A\cap B=A\wedge B\Leftrightarrow A\wedge B\leq A$.)
\end{enumerate}
\vskip 3mm
The set of extended events $\mathcal E_\Omega$ and the set of partitions $\Pi_\Omega$ are, in fact, isomorphic. Let us assume now that $\Omega$ is finite.

\Definition 
\begin{enumerate}[(i)]
	\item For each partition $A=\{A_\alpha|\alpha\in I\}\in\Pi_\Omega$ the associated event $A^{(ev)}$ is defined by
$$A^{(ev)}=\vee\{\sqcup A_\alpha|\alpha\in I\}.$$
(It is clear that this map is an isomorphism.)
  \item For $a:=A^{(ev)},\ b=B^{(ev)},\ A,B\in\Pi_\Omega$ we set
\begin{gather*}
a\wedge b:=(A\wedge B)^{(ev)},\ a\vee b:=(A\vee B)^{(ev)}\\
\neg a:=(\neg A)^{(ev)}
\end{gather*}
\end{enumerate}
Clearly, for each event  $e\in\mathcal E_\Omega$ there exists a unique partition $E\in\Pi_\Omega$ such that
$$e=E^{(ev)}.$$
\vskip 3mm

\Remark Let $a=A^{(ev)},\ b=B^{(ev)}$. Then
\begin{enumerate}[(i)]
	\item $a\vee b$ coincides with the previously introduced operation in the case  when $a,\ b$ are atomic and disjoint
	\item $a\pitchfork b$ iff $A\pitchfork B$
	\item $a$ is a classical (irreducible) iff $A\in\Pi_\Omega^{cl}(\Pi_\Omega^{irr})$
\end{enumerate}
\vskip 3mm

An event can generate the set of events by

\Definition Let $a=A^{(ev)}\in\mathcal E_\Omega$, $A\in\Pi_\Omega$. We set $\bar{a}=\{b\in\mathcal E_\Omega|b\leq a\},\ \bar{A}=\{B\in\Pi_\Omega|B\leq A\}.$ (Evidently $\bar{a}=\{B^{(ev)}|B\in\bar{A}\}$.)

\Definition For $a=A^{(ev)}$ we define $\sqcup a:=\neg\neg a$\\
Then we have
\begin{enumerate}[(i)]
	\item $\sqcup a=\sqcup(\spt A)=(\sqcup A)^{(ev)}$
	\item $\sqcup\sqcup a=\sqcup a$, (i.e. $\sqcup$ is the "closure" operation)
	\item $b=\sqcup a$ iff $b$ is irreducible and $\spt b=\spt a$
	\item $\neg\neg\neg A=\sqcup\neg A=\neg\sqcup A=\neg A$
\end{enumerate}

\vfill\eject
\section{Contexts and universes}

Let us consider the question which events can be observed in a given experiment. It is clear that two incompatible events cannot be simultaneously observed.
\vskip 2mm

For example, let us consider two atomic events $a,b$ which are incompatible $a\not\pitchfork b$. This implies that $a\ne b$ and that $\spt a\cap\spt b\ne\emptyset$.

The condition $a\ne b$ implies that both equalities $\spt a=\spt a\cap\spt b=\spt b$ cannot be true. We can assume that one of them is not true, say $\spt a\cap\spt b\ne\spt b$. Then in both cases, when $a$ happens and when $a$ does not happen, the irreducibility of $b$ will be destroyed.
\vskip 2mm

Thus if $a\not\pitchfork b$, then $a$ and $b$ cannot be simultaneously observed in the same experiment.
\vskip 2mm

We have arrived at the important conclusion, that only mutually compatible events can be observed in a given experiment.
\vskip 2mm

Let us denote the set of all events observable in the experiment $Exp_1$ by
$$\mathcal K=\mathcal K(Exp_1).$$
The set $\mathcal K$, called the context  of $Exp_1$ must have the  following properties
\begin{enumerate}[(i)]
	\item $\pitchfork(\mathcal K)$ i.e. all events in $\mathcal K$ are compatible
	\item $\mathcal K$ is the maximal set of compatible events i.e. for each  event $e\not\in\mathcal K$ there exists $f\in\mathcal K$, such that  $e\not\pitchfork f$.
\end{enumerate}

For each  experiment, its context must be specified and the definition  of experiment's context makes the necessary part of the definition of the experiment.
\vskip 3mm

These arguments leads to  the following basic definition of a concept of a context.

\Definition A subset $\mathcal K\subset\mathcal E_\Omega$ is called a context if $\pitchfork(\mathcal K)$ and if
$$\mathcal K'\supset\mathcal K,\ \pitchfork(\mathcal K')\Rightarrow\mathcal K'=\mathcal K.$$

The set of all contexts in $\mathcal E_\Omega$ is denoted $\Kon_\Omega$
\vskip 3mm
The basic properties of contexts are listed in the following proposition.

\Proposition 
\begin{enumerate}[(i)]
	\item For each context $\mathcal K$ there exists a unique event $u_{\mathcal K}\in\mathcal K$ called the universe of $\mathcal K$ satisfying
	$$\mathcal K=\{e\in\mathcal E_\Omega|e\leq u_{\mathcal K}\}$$
	\item $u_{\mathcal K}$ is the reducible union of atoms from $\mathcal K$, i.e.
	$$u_{\mathcal K}=\vee\{a\in\mathcal K|a\hbox{ is an atom}\}$$
	\item An event $u\in\mathcal K$ is the universe of $\mathcal K$ iff  $\spt u=\Omega$
	\item If $\mathcal K_1,\mathcal K_2\in \Kon_\Omega$, then
	$$\mathcal K_1\ne\mathcal K_2\Leftrightarrow u_{\mathcal K_1}\ne u_{\mathcal K_2}\Leftrightarrow u_{\mathcal K_1}\not\pitchfork u_{\mathcal K_2}$$
\end{enumerate}

\Definition An event $u\in\mathcal E_\Omega$ is a universal event (a universe) iff $\spt u=\Omega$.

The set of all universal events  in $\mathcal E_\Omega$ will be denoted $\Univ_\Omega$.

\Proposition
\begin{enumerate}[(i)]
	\item An event $u\in\mathcal E_\Omega$ is a universe iff there exists a context $\mathcal K$ such that $u=u_{\mathcal K}$.
	\item If $u_1,u_2\in\Univ_\Omega,\ u_1\ne u_2$ then $u_1\not\pitchfork u_2$
	\item The map
	\begin{gather*}
  \Phi:\Kon_\Omega\rightarrow\Univ_\Omega\\
	\mathcal K\mapsto u_{\mathcal K}
	\end{gather*}
	is a 1-1 map onto $\Univ_\Omega$.\\
	The inverse map is given by $u\mapsto\mathcal K_u:=\{e\in\mathcal E_\Omega|e\leq u\}$.
\end{enumerate}
\vskip 3mm
There are two important contexts and universes.

\Definition
\begin{enumerate}[(i)]
	\item the classical context is defined by the classical universe
	$$u_\Omega^{cl}=\vee(\Omega)=e_1\vee\dots\vee e_n.$$
	clearly $\mathcal K_\Omega^{cl}$ contains exactly classical events
	$$\mathcal K_\Omega^{cl}=\{A^{(ev)}|A\in\Pi_\Omega^{(cl)}\}=\{e_{i_1}\vee\dots\vee e_{i_k}\}$$
	\item the irreducible  context is defined by the irreducible universe
	$$u_\Omega^{irr}:=\sqcup(\Omega)=e_1\sqcup\dots\sqcup e_n$$
	and we have
	$$\mathcal K_\Omega^{irr}=\{\Phi,u_\Omega^{irr}\}.$$
\end{enumerate}
\vskip 3mm
Each context has a structure of Boole algebra if the operation of the complement is properly defined

\Definition Let $\mathcal K$ be a context. For each $e\in\mathcal K$ we set
$$\neg_{\mathcal K}e:=\vee\{b\in\mathcal K|b\bot e, b \hbox{ is an atom}\}.$$
\vfill\eject
\Proposition
\begin{enumerate}[(i)]
	\item We have (using the preceding section)
	$$\neg_{\mathcal K}e=(\neg e)\wedge u_{\mathcal K},\ \forall e\in\mathcal K$$
	\item $(\mathcal K,\emptyset,u_{\mathcal K},\wedge,\vee,\neg_{\mathcal K})$ is a Boole algebra.
\end{enumerate}
\vskip 3mm
The concept of context is fundamental in EPT. The description of an experiment means that  the set of observable events is completely specified. I.e. that the context of the experiment is uniquelly determined.
\vskip 2mm

It is not true, that the context of the experiment can be choosen freely.

On the contrary: the experiment must be described in such a way, that this description implies which events are observable. (Physicists usually very clearly describe which events are observable in  a given experiment.)
\vskip 3mm

It is useful to give the general  probability description of the well-known two-slit experiment as a typical example  clarifying the meaning of the context.
\vskip 3mm

\textbf{Example 5.1}(two-slit experiment).\\
Let $n\geq 2$ be fixed and we set
$$\Omega=\{e_{11},e_{21},e_{12},e_{22},\dots,e_{1n},e_{2n}\}=\{e_{ix}|i=1,2, x=1,\dots,n\}.$$
Here $i=1,2$ corresponds to two slits, while $x=1,\dots,n$ correspond to the position on the screen.
\vskip 2mm

There are two typical situations which are characterized by two different contexts.

The first context is given by classical universe
$$u_{\mathcal K_1}=e_{11}\vee e_{21}\vee\dots\vee e_{1n}\vee e_{2n}.$$
$\mathcal K_1$ describes the situation  where the which-way information is available, i.e. when the particle passes through slits in the distinguishable way.

The second context $\mathcal K_2$ is defined by the universe
$$u_{\mathcal K_2}=(e_{11}\sqcup e_{21})\vee\dots\vee(e_{1n}\sqcup e_{2n}).$$
$\mathcal K_2$ describes the situation where the  which-way information is not available, i.e. the particle passes through slits in an in-distinguishable way.

If we observe the particle on the screen at the position $x\in\{1,\dots,n\}$, then in the first experiment we observe the event 
$$e_{1x}\vee e_{2x}$$
while in the second experiment we observe the event
$$e_{1x}\sqcup e_{2x}.$$
(It is clear that different  events can have different probabilities!)
\vskip 2mm

In this way the which-way information enters into physics: through the specification of the experiment's context.
\vskip 2mm

The incompatibility of events $e_{1x}\vee e_{2x}$ , $e_{1x}\sqcup e_{2x}$ can be stated in the following form; in the given experiment  it is impossible simultaneously  to have and not to have the which-way information.
\vskip 2mm

It is completely clear that this incompatibility has purely logical origin based only on the requirement of the logical consistency.
\vskip 2mm

It must be noted that this example is not a correct description of the quantum two-slit  experiment. The role played by the two contexts is only analogical to the situation in QM, so that Example 5.1 describes  the situation in EPT which does not exists in QM. 
\vskip 2mm

QM can be represented in EPT, but this needs more complicated tools (the symplectic structure in EPT) and this will be described later.

\vfill\eject
\section{Relative frequency, extended measures, extended probability  spaces.}

We have introduced contexts as maximal sets of compatible events and we have seen that each context has the structure of Boole algebra. 

It is natural to expect that in each context there is given the standard classical probability theory.

As a first step we specify clearly what is the measurable space associated to $\mathcal K\in \Kon_\Omega$.

We shall denote by $\Omega_{\mathcal K}$ the set of atoms in $\mathcal K$
$$\Omega_{\mathcal K}:=\{a\in\mathcal K| a \hbox{ is an atomic event}\}.$$
Each event $e\in\mathcal K$ can be represented as a subset of $\Omega_\mathcal K$ by the natural association 
$$e^{\mathcal K}:=\{a\in\Omega_\mathcal K|\spt a\subset \spt e\}.$$
Then operation $\wedge,\ \vee,\ \neg_{\mathcal K}$ can be simply represented: for $e,f\in\mathcal K$ we have
\begin{gather*}
(e\wedge f)^{\mathcal K}=e^{\mathcal K}\cap f^{\mathcal K}\\
(e\vee f)^{\mathcal K}=e^{\mathcal K}\cup f^{\mathcal K}\\
(\neg_{\mathcal K}e)^{\mathcal K}=\Omega_{\mathcal K}\backslash e^{\mathcal K}.
\end{gather*}

For a finite set $\Omega$ there exists a canonical algebra of all subsets
$$\mathcal A_{\Omega}:=\{A|A\subset\Omega\}$$
We see that the algebra
$$(\mathcal K,{\bf 0},u_{\mathcal K},\wedge,\vee,\neg_{\mathcal K})$$
is isomorphic to the standard Boole algebra
$$(\mathcal A_{\Omega_{\mathcal K}},\emptyset ,\Omega_{\mathcal K},\cap,\cup,\backslash).$$
\vskip 3mm

The meaning of our approach  requires that in each context there is given a classical probability theory CPT${}_\mathcal K$. There is a natural question how these CPT${}_{\mathcal K_1}$ , CPT${}_{\mathcal K_2}$ are inter-related. This question will be now considered.
\vskip 3mm

We can suppose that for each context $\mathcal K\in\Kon_\Omega$ there exists a measure $\mathbb F_{\mathcal K}$ such that
$$(\Omega_{\mathcal K}, \mathcal A_{\Omega_{\mathcal K}}, \mathbb F_{\mathcal K})$$ will be a Kolmogorov probability space which is a model for the classical probability theory
$$(\mathcal K,{\bf 0},u_{\mathcal K},\wedge,\vee,\neg_{\mathcal K}).$$
There is a question, if there exist some relations between $\mathbb F_{\mathcal K_1}$ and $\mathbb F_{\mathcal K_2}$ for $\mathcal K_1\ne\mathcal K_2$.
\vskip 3mm

The assumption
$$\mathbb F_{\mathcal K_1}(a)=\mathbb F_{\mathcal K_2}(a),\ \forall a\in\mathcal K_1\cap\mathcal K_2$$
is too strong. The weaker assumption requires only that the quotiens of frequences are invariant
$$\frac{\mathbb F_{\mathcal K_1}(a)}{\mathbb F_{\mathcal K_1}(b)}=\frac{\mathbb F_{\mathcal K_2}(a)}{\mathbb F_{\mathcal K_2}(b)},\ \forall a,b\in\mathcal K_1\cap\mathcal K_2$$
It is possible to show that this relation (together with some other technical assumptions) implies the existence of a function
$$P:\mathcal E_\Omega\rightarrow[0,\infty)$$
satisfying
$$\mathbb F_{\mathcal K}(a)=\frac{P(a)}{P(u_{\mathcal K})},\ \forall a\in\mathcal K.$$
The formulation and the proof of this fact is rather long and technical, so that we prefer to postpone this part and to assume  directly the existence of $P$.
\vskip 3mm

There is also another  complication related to the possibility that $P(u_{\mathcal K})=0$.
\vskip 3mm

All this motivates the following definition 

\noindent\textbf{Definition: } Let us consider the function
$$P:\mathcal E_\Omega\rightarrow[0,\infty).$$
\begin{enumerate}[(i)]
	\item A context $\mathcal K\in\Kon_\Omega$ is $P$-regular iff
	$$P(u_{\mathcal K})>0$$
	\item $P$ is an extended measure iff
	$$P_{|\mathcal K}:\mathcal K\rightarrow[0,\infty)$$
	is a measure $\forall\mathcal K\in\Kon_\Omega$
	\item For each $P$-regular context $\mathcal K\in\Kon_\Omega$ we set
	$$\mathbb F_{\mathcal K}(a)=\frac{P(a)}{P(u_{\mathcal K})},\ a\in\mathcal K$$
\end{enumerate}

\Proposition Let the function $P:\mathcal E_\Omega\rightarrow[0,\infty)$ satisfies conditions
\begin{enumerate}[(i)]
	\item if $a_1,\dots,a_s\in\mathcal E_\Omega$ are disjoint atoms i.e. $\bot(a_1,\dots,a_s)$ then
	$$P(a_1,\dots,a_s)=P(a_1)+\dots+P(a_s)$$
	\item $P(\emptyset)=0$\\
\end{enumerate}
Then $P$ is an extended measure.

\Proof Consider the context $\mathcal K,a_1,\dots,a_s\in\mathcal K,\ \bot(a_1,\dots,a_s)$ then $P$ is an additive measure on $\mathcal A_\Omega$ (we assume that $\Omega$ is finite).

\Remark The opposite assertion is also clear: each extended measure satisfies (i) and (ii). If $(a_1,\dots,a_s)$ are disjoint atoms, then surely exists a context $\mathcal K$ such that $a_1,\dots,a_s\in\mathcal K$.

\Proposition Let $P:\mathcal E_\Omega\rightarrow[0,\infty)$ be an extended measure and $\mathcal K$ be $P$-regular context. Then
$$(\Omega_{\mathcal K},\mathcal A_{\Omega_{\mathcal K}},\mathbb F_{\mathcal K})$$
is the Kolmogorov model of CPT, where $\mathbb F_{\mathcal K}$ is defined on $\mathcal A_{\Omega_{\mathcal K}}$ by
$$\mathbb F_{\mathcal K}(e^{\mathcal K})=\mathbb F_{\mathcal K}(e),\ e\in\mathcal K.$$

\Remark If $\mathcal K$ is $P$-irregular, $P(u_{\mathcal K})=0$ then we can assume that $u_{\mathcal K}$ never happens and that irregular contexts may be omitted.
\vskip 3mm

Now we can define the main concept, the extended probability space, which generalizes the Kolmogorov probability space.

\Definition The triple
$$(\Omega,\mathcal E_\Omega, P)$$
is called the extended probability space iff 
\begin{enumerate}[(i)]
	\item $\Omega$ is a (finite) non-empty set - the set of elementary events
	\item $\mathcal E_\Omega$ is the set of extended events
	\item $P:\mathcal E_\Omega\rightarrow[0,\infty)$ is the extended measure
	\item the classical context $\mathcal K^{cl}$ is $P$-regular, i.e. $P(u^{cl})>0$.
\end{enumerate}
\vskip 3mm

\Remark The normalization $P(u^{cl})=1$ is always possible, but it is un-necessary. In fact, the change $P\mapsto k\cdot P,\ k>0$ does not introduce any change in: frequences $\mathbb F_{\mathcal K}$, $P$-regularity, the set of null-events
$$\mathcal N:=\{e\in\mathcal E_\Omega|P(e)=0\}$$
On the other hand, if $\Omega$ is infinite, then already the definition of the classical context is problematic. The best way is to ask only $P(u^{cl})>0$.

\vfill\eject
\section{Quadratic representation of partitions and events}

Partitions have rather complicated structure, in fact, they are sets of subsets. This is two-level structure and it is surely more complicated then the structure of subsets (this is one-level structure).
\vskip 2mm

Fortunately, there exists the canonical representation of a partition  as a subset in the Cartesian product $\Omega^2=\Omega\times\Omega$. 
\vskip 2mm

\Warning In this section we shall consider the general set $\Omega$.
\vskip 2mm

Each partition (general $\Omega$)
$$A=\{A_\alpha|\alpha\in I\}\in\Pi_\Omega$$
canonically defines a relation $R_A$ on $\Omega$ by
$$xR_A y\Leftrightarrow\exists\alpha\in I\hbox{ such that }x,y\in A_\alpha,\ x,y\in\Omega$$
(i.e. $x$ and $y$ are inter-related iff they  belong to the same part of $A$).
\vskip 2mm

\Remark Let us note that the relation $R_A$ is symmetric, i.e. $xR_Ay\Rightarrow yR_Ax$.
\vskip 2mm

Each relation $R$ on $\Omega$ defines canonically the subset $\tilde{R}$ of $\Omega\times\Omega$ by
$$\tilde{R}:=\{(x,y)\in\Omega\times\Omega|xRy\}.$$

In fact, this is the  set-theoretical representation of $R$. Putting both representations together, we obtain

\Definition Let $\Omega$ be an \textbf{arbitrary} not-empty set.
\begin{enumerate}[(i)]
	\item For $A\subset \Omega$ we set
	$$A^2:=A\times A:=\{(x,y)\in\Omega\time\Omega|x,y\in A\}$$
	\item The subset $R\subset \Omega^2=\Omega\times\Omega$ is symmetric iff
	$$(x,y)\in R\Rightarrow (y,x)\in R$$
	the set of all symmetric $R$'s is denoted by $\Sym_{\Omega^2}$
	\item We shall say that $R\subset\Omega^2$ is symmetric transitive iff $R$ is symmetric and
	$$xRy,\ yRz\Rightarrow xRz.$$
	The set of all $R\subset\Omega^2$ which are symmetric and transitive will be denoted $ST_{\Omega^2}$
	and these sets will be called $ST$-sets.
	\item For $R\in ST_{\Omega^2}$, the support or $R$ is given by
	$$\spt R:=\{x\in\Omega | (x,x)\in R\}$$
	We also set
	$$\diag \Omega^2=\{(x,x)\in\Omega^2|x\in\Omega\}$$
	\item for $R,\ S\in ST_{\Omega^2}$ we shall define operations
\begin{eqnarray*}
                  \neg R&\!\!:=&\!\!(\Omega\backslash\spt R)^2,\\
               R\wedge S&\!\!:=&\!\!R\cap S,\\
                 R\vee S&\!\!:=&\!\!(R\cap\neg S)\cup(R\cap S)\cup(\neg R\cap S),\\
           R\backslash S&\!\!:=&\!\!R\cap\neg S,\\
R_1\sqcup\dots\sqcup R_s&\!\!=&\!\!\sqcup(R_1,\dots,R_s):=(\spt R_1\cup\dots\cup R_s)^2,\\
                \sqcup R&\!\!:=&\!\!(\spt R)^2
\end{eqnarray*}
  \item We shall use the following definitions\\
	$R$ is classical iff $R\subset\diag\Omega^2$\\
	$R\leq S$ iff $R=R\cap S$\\
	$R\pitchfork S$ iff $[R\cap S\leq R\hbox{ and } R\cap S\leq S]$
	\item $R\in ST_{\Omega^2}$ is called the quadratic set iff there exists $A\subset\Omega$ such that $R=A^2$
	\item $R$ is a universe iff $\spt R=\Omega$.
\end{enumerate}
\vskip 2mm

Basic properties of $ST$-sets are described in
\Proposition Let $R,S\in ST_{\Omega^2}$. Then
\begin{enumerate}[(i)]
	\item $\neg R, R\wedge S, R\vee S, R\backslash S, \sqcup R\in ST_{\Omega^2}$
	\item $(x,y)\in R\Rightarrow (x,x), (y,y)\in R$
	\item $R\leq S$ iff $R=S\cap(\spt R)^2$
	\item $R\pitchfork S$ iff $R\cap(\spt S)^2=S\cap (\spt R)^2$
	\item $\sqcup R=\neg\neg R$
\end{enumerate}
\vskip 2mm

Now we shall define the fundamental connection between partitions an $ST$-sets.

\Definition For each $A\in\Pi_\Omega$,

$$A=\{A_\alpha|\alpha\in I\}$$
we define its quadratic representation by
$$A^Q:=\bigcup_{\alpha\in I}(A_\alpha\times A_\alpha)\subset\Omega\times\Omega$$
\vskip 3mm
Here are the basic properties of this representation

\vfill\eject
\Proposition
\begin{enumerate}[(i)]
	\item $A^Q\in ST_{\Omega^2}$ for each $A\in\Pi_\Omega$
	\item For each $R\in ST_{\Omega^2}$ there exists exactly one $A\in\Pi_\Omega\hbox{ such that }A^Q=R.$\\
	Such $A$ can be defined by
	$$A=\{[x]|x\in\spt R\}$$
	where
	$$[x]=\{y\in\Omega|(x,y)\in R\}$$
	and $[x]$ and $[y]$ are identified if $[x]=[y]$.
	\item The quadratic representation is an isomorphism:
	\begin{enumerate}[]
		\item $(\neg A)^Q=\neg A^Q$,
		\item $(A\wedge B)^Q=A^Q\wedge B^Q$,
		\item $(A\vee B)^Q=A^Q\vee B^Q$,
		\item $A\leq B\hbox{ iff } A^Q\leq B^Q$,
		\item $A\pitchfork B\hbox{ iff }A^Q\pitchfork B^Q$
		\item $A\in\Pi_\Omega^{irr}\Leftrightarrow A^Q$ is quadratic,
		\item $A\in\Pi_\Omega^{cl}$ iff $A^Q$ is classical
	\end{enumerate}
\end{enumerate}

\Proof The proof is simple. Using the quadratic representation also the proof of the preceding proposition is simple.
\vskip 3mm

Using the representation of (extended) events by partitions and the quadratic representation of partitions by $ST$-sets in $\Omega^2$ we have two isomorphism
$$A^{(ev)}\in\mathcal E_\Omega\leftrightarrow A\in\Pi_\Omega\leftrightarrow A^Q\in ST_{\Omega^2}.$$

The composition gives the quadratic representation of events by $ST$-sets
$$A^{(ev)}\in\mathcal E_\Omega\leftrightarrow A^Q\in ST_{\Omega^2}.$$
\vskip 3mm

There are interesting and very important set-theoretical relationes in $ST_{\Omega^2}$. To describe these relations we need some new concepts.
\vskip 2mm

Let $X$ be a non-empty set of any cardinality.

Let $\mathcal A$ be set of subsets of $X$.

The extended algebra $\mathcal A^{\mathbb Z}$ is defined as the set of all finite $\mathbb Z$-valued linear combination of characteristic functions of sets from $\mathcal A$: if $f:X\rightarrow\mathbb Z$ then $f\in\mathcal A^{\mathbb Z}$ iff
$\exists c_1,\dots,c_s\in\mathbb Z,\ \exists A_1,\dots,A_s\in\mathcal A$ such that
$$f=c_1\chi(A_1)+\dots+c_s\chi(A_s)$$
where the characteristic function of $A$ is defined by
$$\chi(A;x)=1\hbox{ iff } x\in A\hbox{ and }\chi(A;x)=0\hbox{ iff }x\not\in A.$$
\vskip 3mm

It is clear that $\mathcal A^{\mathbb Z}$ is the additive closure of the set of characteristic functions. (If $\mathcal A$ is not explicitely defined, we shall assume that $\mathcal A$ is the algebra of all subsets of $X, \mathcal A=\mathcal A_X$.)
\vskip 3mm

Now we are able to state and prove the basic properties of quadratic sets in $ST_{\Omega^2}$

\Proposition\textbf{1.} Let $R_1,\dots,R_s$ be quadratic $ST$-sets in $\Omega^2$ which are disjoint, $\bot(R_1,\dots,R_s)$. Then
$$\chi(R_1\sqcup\dots \sqcup R_s)=\sum_{1=i<j}^s\chi(R_i\sqcup R_j)-(s-2)\sum_{i=1}^s\chi(R_i).$$

\Remark The important case is $s=3$ and then we have
$$\chi(R_1\sqcup R_2\sqcup R_3)=\chi(R_1\sqcup R_2)+\chi(R_1\sqcup R_3)+\chi(R_2\sqcup R_3)-[\chi(R_1)+\chi(R_2)+\chi(R_3)]$$
This relation can be expressed in the set-theoretical form
$$(R_1\sqcup R_2\sqcup R_3)\backslash(R_1\vee R_2\vee R_3)=[(R_1\sqcup R_2)\backslash(R_1\vee R_2)]\cup[(R_1\sqcup R_3)\backslash(R_1\vee R_3)]\cup[(R_2\sqcup R_3)\backslash(R_2\vee R_3)].$$

\Remark It must be stressed that this relation is the set-theoretical relation which does not contain any relation to any measure. In fact, there is no measure mentioned in the statement of Propositon.
\vskip 3mm

\Proof There exist $A_1,\dots,A_s\subset\Omega$ disjoint such that $R_1=A_1^2,\dots, R_s=A_s^2$. Then we have
$$\chi((A_1\cup\dots\cup A_s)^2)=\chi(\bigcup_{i,j}A_i\times A_j)=\sum_{i,j}\chi(A_i\times A_j)$$
and we have on the other hand for each $i\ne j$
$$\chi((A_i\cup A_j)^2)=\chi(A_i^2\cup A_j^2\cup(A_i\times A_j)\cup(A_j\times A_i))=\chi(A_i^2)+\chi(A_j^2)+\chi(A_i\times A_j)+\chi(A_j\times A_i)$$
and then
$$\sum_{i\ne j}^s((A_i\cup A_j)^2)=(s-1)\sum_i\chi(A_i)^2+(s-1)\sum_j\mathcal  X(A_j^2)+2\cdot\sum_{i\ne j}\chi(A_i\times A_j)$$
and then clearly (since the left hand side is symmetric in $i,j$)
$$\sum_{i<j}\chi((A_i\cup A_j)^2)=(s-1)\sum\chi(A_i^2)+\sum_{i\ne j}\chi(A_i\times A_j)=(s-2)\sum\chi(A_i^2)+\sum_{i,j}\chi(A_i\times A_j).$$
Then we obtain
$$\chi((A_1\cup\dots\cup A_s)^2)=\sum_{i<j}\mathcal((A_i\cup A_j)^2)-(s-2)\sum_{i=1}\chi(A_i^2).\hbox to \hsize{\hfill }$$

\Remark Let $A=\{x_1,\dots, x_s\}\subset\Omega$. Then
$$(+)\hbox to 15mm{}\chi(A^2)=\sum_{1\leq k < l}^s\chi(\{x_k,x_l\}^2)-(s-2)\sum_{k=1}^s\chi(\{x_k\}^2).$$

Quadratic $ST$-set $B$ is called a dyadic atom iff there exist $x,y\in\Omega, x\ne y$ such that
$$B=\{x,y\}^2=\{(x,x),(y,y),(x,y),(y,x)\}.$$
Thus each finite set in $ST_{\Omega^2}$ can be expressed using only classical and dyadic atoms. This is especially important in the  case when $\Omega$ is finite. Then characteristic function of each $R\in ST_{\Omega^2}$ can be written  as a linear combination of characteristic functions of classical and dyadic atoms: $\{x\}^2, \{x,y\}^2,\ x,y\in\Omega,\ x\ne y$.
\vskip 3mm
At the end we can say that events in $\mathcal E_\Omega$ can be truth-fully represented as $ST$-sets in $\Omega^2$ by the quadratic representation.
\vskip 3mm

Now we shall generalize these results to the $\mathbb Z$-valued functions defined above. We shall consider functions from $\mathcal A^{\mathbb Z}$, where $\mathcal A$ is the algebra of subsets of $\Omega,\ \mathcal A=\mathcal A_\Omega=\{A|A\subset\Omega\}$ so that $\mathcal A^{\mathbb Z}$ is the space of $\mathbb Z$-valued functions on $\Omega$. Functions on $\Omega^2$ can be constructed by the tensorial product. We shall use the following definition.

\Definition Let $f,g\in\mathcal A^{\mathbb Z}$. Then we denote by $f\otimes g$ the following functions on $\Omega^2$
$$f\otimes g(x,y):=f(x)\cdot g(y),\ (x,y)\in\Omega^2.$$
We shall denote by $f^{\otimes 2}=f\otimes f$ the function
$$f^{\otimes 2}(x,y):=f(x)f(y),\ (x,y)\in\Omega^2.$$
\vskip 3mm
Then the proposition above can be generalized.

\Proposition\textbf{2} Let $f_1,\dots,f_s\in\mathcal A^{\mathbb Z}$. Then
$$\left(\sum_{i=2}^s f_i\right)^{\otimes 2}=\sum_{i<j}(f_i+f_j)^{\otimes 2}-(s-2)\sum f_i^{\otimes 2}.$$

\Proof Proof is the same as above. We have
\begin{enumerate}[]
	\item $\left(\sum_i f_i\right)^{\otimes 2}=\sum_{i,j} f_i\otimes f_j$ (since $\otimes$ is bi-linear),
	\item $\left(f_i+f_j\right)^{\otimes 2}=f_i^{\otimes 2}+f_j^{\otimes 2}+f_i\otimes f_j+f_j\otimes f_i$
	\item and then
	\item $\sum_{i\ne j}(f_i+f_j)^{\otimes 2}=(s-1)\sum f_i^{\otimes 2}+(s-1)\sum f_j^{\otimes 2}+2\sum_{i\ne j}f_i\times f_j$.
	\item At the end we obtain
	\begin{eqnarray*}
	  \sum_{i<j}(f_i+f_j)^{\otimes 2}&=&(s-1)\sum f_i^{\otimes 2}+\sum_{i\ne j}f_i\otimes f_j=\\
		                               &=&(s-2)\sum f_i^{\otimes 2}+\sum_{i,j}f_i\otimes f_j=\\
																 	 &=&(s-2)\sum f_i^{\otimes}+\left(\sum f_i\right)^{\otimes 2}.\hbox to 2cm{\hfill $\square$}
	\end{eqnarray*}
\end{enumerate}

As a consequence we obtain in the case $s=3$

\Proposition\textbf{3}
$$(f_1+f_2+f_3)^{\otimes 2}=(f_1+f_2)^{\otimes 2}+(f_1+f_3)^{\otimes 2}+(f_2+f_3)^{\otimes 2}-(f_1^{\otimes 2}+f_2^{\otimes 2}+f_3^{\otimes 2}).$$
\vskip 3mm

This relation can be reformulated in the following form

\Proposition\textbf{4} Let $f_1,f_2,f_3\in\mathcal A^{\mathbb Z}$ and let $g=f_2+f_3$. Then
$$(f_1+g)^{\otimes  2}-f_1^{\otimes  2}-g^{\otimes  2}=\left[(f_1+f_2)^{\otimes  2}-f_1^{\otimes  2}=f_2^{\otimes  2}\right]+\left[(f_1+f_3)^{\otimes  2}-f_1^{\otimes  2}-f_3^{\otimes  2}\right].$$
This means the linearity of the form
$$(f+g)^{\otimes  2}-f^{\otimes  2}-g^{\otimes  2}$$
in $g$.

\Proof This is a direct consequence of \textbf{Proposition 3.}. Using it we obtain
\begin{eqnarray*}
(f_1+g)^{\otimes  2}-f_1^{\otimes  2}-g^{\otimes  2}&=&(f_1+f_2+f_3)^{\otimes  2}-f_1^{\otimes  2}-(f_2+f_3)^{\otimes  2}=\\
                                &=&\left[(f_1+f_2)^{\otimes  2}-f_1^{\otimes  2}-f_2^{\otimes  2}\right]+\left[(f_1+f_3)^{\otimes  2}-f_1^{\otimes  2}-f_3^{\otimes  2}\right].
\end{eqnarray*}

Functions used in \textbf{Proposition 2}-\textbf{Proposition 4} are from the set of symmetric $\mathbb Z$-valued functions on $\Omega^2$.

\Definition We denote
$$\Sym_{\Omega^2}^{\mathbb Z}=\{f:\Omega^2\rightarrow\mathbb Z|f(x,y)=f(y,x),\ \forall x,y\in\Omega\}.$$
\vskip 3mm

We shall assume that $\Omega$ is finite, $\Omega=\{e_1,\dots,e_n\}$. We want to make clear what are all linear dependeces among functions from $\Sym_{\Omega^2}^{\mathbb Z}$. It is important to define the canonical  bases in $\Sym_{\Omega^2}^{\mathbb Z}$.

\Definition Let us denote for $(x,y)\in\Omega^2$
\begin{enumerate}[]
	\item $\delta_{xy}=\chi((x,y))$,
	\item $h_{xy}=\chi(\{x,y\}^2)=\delta_{xx}+\delta_{yy}+\delta_{xy}+\delta_{yx}$,
	\item $g_x=\delta_{xx}$.
\end{enumerate}

\Proposition
\begin{enumerate}[(i)]
	\item The set of functions
	$$\mathcal B_{\Omega^2}:=\{h_{xy}|1\leq x<y\leq n\}\cup\{g_x|1\leq x\leq n\}$$
	forms the $\mathbb Z$-bases of $\Sym_{\Omega^2}^{\mathbb Z}$, i.e. each symmetric $\mathbb Z$-valued functions can be expressed (in a unique way) as a $\mathbb Z$-valued linear combination of function from $\mathcal B_{\Omega^2}$.
	\item If we have
	$$f=\sum_{k,l}f_{kl}\delta_{kl}\in\Sym_{\Omega^2}^{\mathbb Z},\hbox{ i.e } f_{kl}=f_{lk}\in\mathbb Z$$
	then
$$f=\sum_{x<y}f_{xy}h_{xy}+\sum_{x}f_{xx} g_x-\sum_{y\ne z}f_{yz} g_y.$$
\item The characteristic function of any atom $A^2, A\subset\Omega$ can be written as a $\mathbb Z$-valued linear combination of functions from $\mathcal B_{\Omega^2}$.
\end{enumerate}

\Proof
\begin{enumerate}[(i)]
	\item We shall use substitutions $(k\ne l)$
	\begin{enumerate}[]
		\item $\delta_{kl}+\delta_{lk}=h_{kl}-g_k-g_l$
		\item $\delta_{kk}=g_k$;
	\end{enumerate}
	the independence of functions in $\mathcal B_{\Omega^2}$ is clear:\\
	the standard basis has the form
	$$\{\delta_{rs}+\delta_{sr}|r<s\}\cup\{\delta_{rr}\}$$
	and this is equivalent to $\{h_{rs}|r<s\}\cup\{g_r\}$
	\item it follows by the explicite calculation using
	$$f_{xy}=f_{yx}.$$
	\item the characteristic function of any atom can be written as a $\mathbb Z$-linear combination of characteristic functions of classical and dyadic atoms.
\end{enumerate} $\square$
\vskip 3mm

Now we shall consider the following question: what are all $\mathbb Z$-valued linear dependences in $\mathcal E_\Omega$?
We shall start with the basic set of dependences
\begin{enumerate}[(*)]
	\item $\chi(A^2)=\sum(\{\chi(e_k,e_l)\}^2)|k<l, e_k, e_l\in A\}-(s-2)\sum\{\chi(\{e_k\}^2)|e_k\in A\}.$
\end{enumerate}
\vskip 3mm

We have the following proposition

\textbf{Proposition.} Let $A_1,\dots,A_m\subset\Omega$ are mutually different sets in $\Omega$. Let us assume that there exist integers $c_1,\dots,c_m$ such that
$$c_1\chi(A_1^2)+\dots+c_m\chi(A_m^2)=0.$$

Then this relation can be obtained as a $\mathbb Z$-valued linear combinations of the relations (*).

\Proof If all atoms $A_1^2,\dots,A_m^2$ are classical or dyadic, then this  linear dependence contradicts to the independence of  the basis $\mathcal B_{\Omega^2}$.
Let $A_1$ is such that $|A_1|\geq 3$. We can express all $\chi(A_2),\dots,\chi(A_m)$ using (*). Then $\chi(A_1)$ will be written as a combination of classical and dyadic atoms. The resulting expression must be an integer multiple of (*). Transforming all $\chi(A_2^2),\dots,\chi(A_0^2)$ back we obtain the conclusion.
\vskip 3mm

\Conclusions
\begin{enumerate}[(i)]
	\item Characteristic function of any atom $A^2, A\in\Omega$ can be written as a $\mathbb Z$-linear combination of characteristic functions of classical and dyadic atoms.
	\item Each symmetric $\mathbb Z$-valued function on $\Omega^2$ can be expressed in the same way.
\end{enumerate}
\vskip 3mm

\textbf{Example 7.1} The example 5.1 has the following form in the quadratic representation\\
\begin{enumerate}[]
	\item $\Omega=\{e_{11},e_{21},\dots,e_{1n},e_{2n}\}$,
	\item $e_{ix}^Q=\{(e_{ix},e_{ix})\}\subset\Omega^2,\ i=1,2,\ x=1,\dots,n$,
	\item $(e_{1x}\vee e_{2x})^Q=\{(e_{1x},e_{1x}),(e_{2x},e_{2x})\}\subset\Omega^2$,
	\item $(e_{1x}\sqcup e_{2x})^Q=\{(e_{1x},e_{1x}),(e_{2x},e_{2x}),(e_{1x},e_{2x}),(e_{2x},e_{1x})\}$,
	\item $(u^{cl})^Q=\diag \Omega^2=\{(e_{11},e_{11}),(e_{21},e_{21}),\dots,(e_{1n},e_{1n}),(e_{2n},e_{2n})\}$
	\item $(u_{\mathcal K_2})^Q=\{e_{11},e_{21}\}^2\cup\dots\cup\{e_{1n},e_{2n}\}^2$
	\item $=\{(e_{11},e_{11}),(e_{21},e_{21}),(e_{11},e_{21}),(e_{21},e_{1}),\dots,(e_{1n},e_{1n}),(e_{2n},e_{2n}),(e_{1n},e_{2n}),(e_{2n},e_{1n})\}$
\end{enumerate}

\vfill\eject
\section{Quadratic representation of the extended probability measure}

Here we shall suppose that $\Omega$ is finite set.
\vskip 3mm

The extended probability measure is the function
$$P:ST_{\Omega^2}\rightarrow[0,\infty)$$
such that if $A_1,\dots,A_s$ are disjoint subset of $\Omega$, then
$$P(A_1^2\cup\dots\cup A_s^2)=P(A_1^2)+\dots+P(A_s^2),$$
where any event
$$\vee_{\alpha\in I}(\sqcup A_\alpha)\in\mathcal E_{\Omega^2}$$
is represented by
$$\bigcup_{\alpha\in I}A_\alpha^2\in ST_{\Omega ^2}.$$
\vskip 3mm

Following the long tradition, we shall consider sets in $ST_{\Omega^2}$ as events, but the isomorphisms
$$\vee_\alpha(\sqcup A_\alpha)\in\mathcal E_\Omega\leftrightarrow\{A_\alpha|\alpha\in I\}\in\Pi_\Omega\leftrightarrow\bigcup_\alpha A_\alpha^2\in ST_{\Omega^2}$$
will always be understood. (It is clear that an event and a subset of $\Omega^2$ are two different things, but in CPT the situation is similar: a classical event and a subset of $\Omega$ are also different things.)
\vskip 3mm
It is assumed in CPT that the probability should be additive with respect to the disjoint union of subsets. Partitions are not subsets, so that the concept of the additivity cannot be directly applied to partitions.
\vskip 3mm

But partitions have the canonical quadratic representation in $ST_{\Omega^2}$ as subsets of $\Omega^2$
$$\{A_{\alpha}|\alpha\in I\}\leftrightarrow\bigcup_{\alpha\in I}A_\alpha^2\subset\Omega^2.$$
We shall require $P$ to be a homomorphism with respect to additivity structure which already exists in $ST_{\Omega^2}$, i.e. $P$ has to be an additivity homomorphism from $ST_{\Omega^2}\ \into\ \mathbb  R$.
\vskip 3mm

In particular we require that $P$ has to be a homomorphism with respect to linear relations expressed in formulas (+) from the preceeding section.

\Definition The extended probability  measure $P$ is called the quadratic probability measure iff  for each subset $A=\{x_1,\dots,x_s\}\in\Omega$ we have
$$P(A^2)=\sum_{1\leq i<j}^s P(\{x_i,x_j\}^2)-(s-2)\sum_{i=1}^s P(\{x_i\}^2).$$
\vskip 3mm
It is clear that it is sufficient to know the quadratic probability measure only on classical $\{x_i\}^2$ and dyadic $\{x_i,x_j\}^2,i<j$ atoms.
\vskip 3mm

This suggests the following definition of the probability distribution corresponding to $P$.

\Definition

\begin{enumerate}[(i)]
	\item Let $P$ be a quadratic probability measure on $\Omega^2$ ($\Omega$ finite!). The probability distribution corresponding to $P$ is the function
	$${\bf p}={\bf p}_P:\Omega^2\rightarrow\mathbb R$$
	defined by
	$${\bf p}(x,x)=P(\{x\}^2),\ x\in\Omega$$
	$${\bf p}(x,y)=\frac{1}{2}[P(\{x,y\}^2)-P(\{x\}^2)-P(\{y\}^2)],\ x,y\in\Omega,\ x\ne y$$
	\item In general, the function 
	$$f:\Omega^2\rightarrow\mathbb R$$
	is called the quadratic probability distribution iff $f$ is symmetric, i.e. $f(x,y)=f(y,x),\ \forall x,y\in\Omega$.
\end{enumerate}
\Remark We see immediately that for $x\ne y$
$$P(\{x,y\}^2)=P(\{x\}^2)+P(\{y\}^2)+2P(x,y)={\bf p}(x,x)+{\bf p}(y,y)+{\bf p}(x,y)+{\bf p}(y,x).$$
\vskip 3mm

The following proposition is the generalization of this simple formula.
\Proposition Let $P$ be a quadratic probability  measure on $\Omega^2$ ($\Omega$ finite!) and let ${\bf p}:\Omega^2\rightarrow\mathbb R$ is the corresponding probability distribution. Then for each $A=\{x_1,\dots,x_s\}\in\Omega$ we have
$$P(A^2)=\sum_{i,j}{\bf p}(x_i,x_j)=\sum\{{\bf p}(x,y)|(x,y)\in A^2\}.$$

\Proof By the definition of $P$ and $x_i$ we have
\begin{eqnarray*}
P(\{x_1,\dots,x_s\}^2)&=&\sum_{i<j}P(\{x_i,x_j\}^2)-(s-2)\sum_{i}P(\{x_i\}^2)=\\
                      &=&\sum_{i<j}[{\bf p}(x_i,x_i)+{\bf p}(x_j,x_j)+{\bf p}(x_i,x_j)+{\bf p}(x_j,x_i)]-(s-2)\sum {\bf p}(x_i,x_i)=\\
											&=&\sum_{i\ne j}{\bf p}(x_i,x_i)+\sum_{i\ne j}(x_i,x_j)-(s-2)\sum {\bf p}(x_i,x_i)\\
											&=&(s-1)\sum_i{\bf p}(x_i,x_i)+\sum_{i,j}{\bf p}(x_i,x_j)-\sum_i{\bf p}(x_i,x_i)-(s-2)\sum {\bf p}(x_i,x_i)\\
			                &=&\sum_{i,j}{\bf p}(x_i,x_j)\hbox to 1cm{}\square\\
\end{eqnarray*}
\vskip 3mm
It is clear that the probability distribution ${\bf p}:\Omega^2\rightarrow\mathbb R$ defines a measure on $\Omega^2$.

\Definition Let $P$ be quadratic probability measure and let ${\bf p}:\Omega^2\rightarrow\mathbb R$ is the corresponding probability distribution. We shall define the signed measure
$$\lambda=\lambda_{\bf p}=\lambda_P$$
on $\Omega^2$ by
$$\lambda(A):=\sum\{{\bf p}(x,y)|(x,y)\in A\},\ A\subset\Omega^2.$$
\vskip 3mm
The measure $\lambda$ have the following properties
\Properties Let $P,{\bf p}$ and $\lambda$ are as above. Then
\begin{enumerate}[(i)]
	\item $\lambda$ is a signed measure on $\Omega^2$ and ${\bf p}$ is its probability density
	\item $\lambda$ is symmetric in the sense that
	$$\lambda(A\times B)=\lambda(B\times A),\ \forall A,B\subset\Omega;$$
	in particular
	$$\lambda((x,y))=\lambda((y,x)),\ \forall x,y\in\Omega,\ x\ne y$$
	\item $P$ coincides with $\lambda$ on $ST_{\Omega^2}$, in particular
	$$P(A^2)=\lambda(A^2),\ \forall A\subset\Omega.$$
\end{enumerate}
\vskip 3mm

\Remarks
\begin{enumerate}[(i)]
	\item The measure $\lambda=\lambda_{\bf p}$ can be defined using $P$ instead of ${\bf p}$ by the following formulas
	\begin{enumerate}[(a)]
		\item $\lambda(\{(x,y),(y,x)\})=P(\{x,y\}^2)-P(\{x\}^2)-P(\{y\}^2),\ x\ne y$
		\item $\lambda(\{x\}^2)=P(\{x\}^2)$
		\item $\lambda(\{x,y\})=\lambda(\{y,x\})$.
	\end{enumerate}
  In fact  (a) and (c) imply that
	$$\lambda((x,y))=\frac{1}{2}[P(\{x,y\}^2)-P(\{x\}^2)-P(\{y\}^2)].$$
	\item From (i) it is clear that there exists exactly one measure $\lambda$ such that $\lambda$  is symmetric and coincides with $P$  on quadratic sets $A^2,\ A\in\Omega$.
	\item It is clear that the algebras of characteristic functions satisfy
	$$ST_{\Omega^2}^{\mathbb Z}=\Sym_{\Omega^2}^{\mathbb Z}$$
	and that $P$  can be extended  $\mathbb Z$ -linearly (in a standard way) from $ST_{\Omega^2}$ onto $ST_{\Omega^2}^{\mathbb Z}$ and  that this extension coincides with $\lambda_P$.
\end{enumerate}
	\vskip 3mm
	
	In what follows we shall make a specific requirement on the positivity of $P$ and $\lambda$.
	
	If $A\subset\Omega$ then we have
	$$0\leq P(A^2)=\int\chi(A^2)d\lambda=\int\chi(A;x)\chi(A;y)d\lambda(x,y).$$
	This may be reformulated as
	$$\int_{\Omega^2} f(x)f(y)d\lambda(x,y)\geq 0$$
	for each $f:\Omega\rightarrow\mathbb R$ such that $f=\chi(A)$ for some $A\subset \Omega$
	\vskip 3mm
	It is useful to consider  the stronger positivity condition with arbitrary $f$'s.
	
	\Definition The quadratic probability  measure
	$P:ST_{\Omega^2}\rightarrow[0,\infty)$ is strongly  positive iff
	$$\int_{\Omega^2}f(x)f(y)d\lambda	_P(x,y)\geq 0$$
	for each  function $f:\Omega\rightarrow\mathbb R$.
	\vskip 3mm
	
	\Remarks The condition of the strong positivity can be formulated in many equivalent ways:
	\begin{enumerate}[(i)]
		\item For each $A_1,\dots,A_s$ disjoint subsets of $\Omega$ the matrix
		$$\left(\lambda(A_i\times A_j)\right)_{i,j=1}^s$$
		is positive semi-definite.
		\item The equivalent formulation using only $P$  is the following. Let $A_1,\dots,A_s$ are disjoint subsets of $\Omega$.\\
		Let us define
		\begin{enumerate}[]
			\item $a_{i,j}=P((A_i\cup A_j)^2)-P(A_i^2)-P(A_j^2),\ i\ne j$
			\item $a_{i,i}=2\cdot P(A_i^2)$
		\end{enumerate}
		and it is required that the matrix$(a_{i,j})$ is positive semi-definite.
		\item The probability distribution ${\bf p}:\Omega^2\rightarrow\mathbb R$, $\Omega=\{e_1,\dots,e_n\}$ is such that  the matrix
		$$(p(e_i,e_j))_{i,j=1}^n$$
		is positive semi-definite.
\end{enumerate}
\vskip 3mm
Using $\lambda_P$, the Proposition 1 from Section 7 can be transformed into the property  of the quadratic probability measure.
	
\Proposition Let $A_1,\dots,A_s\subset\Omega$ are disjoint and let $P$ be a quadratic probability measure. Then
$$P((A_1\cup\dots\cup A_s)^2)=\sum_{1=i<j} P((A_i\cup A_j)^2)-(s-2)\sum_{i=1}^sP(A_i^2).$$
\Proof For any $A\subset\Omega$
$$P(A^2)=\lambda(A^2)=\int\chi(A^2)d\lambda$$
From Propositions 1. sect 7 we obtain
$$\int\chi((A_1\cup\dots\cup A_s)^2)d\lambda=\sum_{i<j}\int\chi((A_i\cup A_j)^2)d\lambda-(s-2)\sum_i\int\chi(A_1^2)d\lambda\ \ \square$$

\vfill\eject
\section{Quadratic probability space}

We shall use some standard measure-theoretical concepts.

\Definition Let $\Omega$ be any non-empty set and let $\mathcal A$ be a $\sigma$-algebra of subsets of $\Omega$.
\begin{enumerate}[(i)]
	\item A measure $\nu:\mathcal A\rightarrow [0,\infty]$ is $\sigma$-finite iff there exists a sequence $A_1,A_2,\dots\in\mathcal A$ such that $A_1\subset A_2\subset\dots,\ \cup A_i=\Omega$ and $\nu(A_i)<\infty,\ \forall i$.
	\item The $\sigma$-algebra $\overline{\mathcal A\times\mathcal A}$ in $\Omega^2$ is the smallest $\sigma$-algebra in $\Omega^2$ containing
	$$\mathcal A\times\mathcal A:=\{A_1\times A_2\subset\Omega^2|A_1,A_2\in\mathcal A\}.$$
	\item The measure $\nu\times\nu$ on $\overline{\mathcal A\times\mathcal A}$ is the unique measure on $\overline{\mathcal A\times\mathcal A}$ satisfying
	$$\nu\times\nu(A_1\times A_2)=\nu(A_1)\cdot\nu(A_2),\ \forall A_1,A_2\in\mathcal A.$$
	\item The signed measure
	$$\lambda:\overline{\mathcal A\times\mathcal A}\rightarrow\mathbb R$$
	is symmetric iff
	$$\lambda(A_1\times A_2)=\lambda(A_2\times A_1),\ \forall A_1, A_2\in\mathcal A$$
	\item If $\nu:\mathcal A\rightarrow[0,\infty]$ is a measure and $f:\Omega\rightarrow\mathbb R$ is a $\nu$-integrable function, then the signed measure $\nu\lefthalfcup f$ is defined by
	$$\nu\lefthalfcup f(A):=\int_A fd\nu.$$
	(Then clearly $f$ is a Radon-Nikodym derivative $f=d(\nu\lefthalfcup f)/d\nu$.)
\end{enumerate}
\vskip 3mm

On the basis of considerations presented in the preceeding section it is natural to introduce our central  concept: the quadratic probability space.

\Definition Quadratic probability space is the triple
$$(\Omega^2,\mathcal E,P)\hbox{ where}$$
\begin{enumerate}[(i)]
	\item $\Omega^2=\Omega\times\Omega$, $\Omega$ is the non-empty set of elementary events
	\item There exists a $\sigma$-algebra $\mathcal A$ on $\Omega$ such that
	$$\mathcal E=ST_{\Omega^2}\cap (\overline{\mathcal A\times\mathcal A}).$$
	\item $P$ is the function
	$$P:\mathcal E\rightarrow[0,\infty)$$
	such that there exists a symmetric signed measure $\lambda$ on $\overline{\mathcal A\times\mathcal A}$ satisfying
	$$P(A)=\lambda(A),\ \forall A\in\mathcal E.$$
	\item $P$ is strongly positive in the sense that
	$$\int_{\Omega^2}f(x)f(y)d\lambda(x,y)\geq 0$$
	for each bounded $\mathcal A$-measurable function $f:\Omega\rightarrow\mathbb R$.\\
	(We shall show below that $\lambda$ is uniquelly determined by $P,$ i.e. $\lambda=\lambda_P$.)
	\item There exists at least one $P$-regular context $\mathcal K$, $P(u_{\mathcal K})>0.$
\end{enumerate}

\Remarks
\begin{enumerate}[(i)]
	\item It is clear that the $\sigma$-algebra $\mathcal A$ is uniquelly determined by $\mathcal E:\mathcal A=\{A\subset\Omega|A^2\in\mathcal E\}$ thus (ii) is the condition on $\mathcal E$.
	\item The signed measure $\lambda$ (if it exists) is uniquelly determined by $P$.
	\begin{enumerate}[(a)]
		\item If $A\cap B=\emptyset$ then
		$$2\cdot\lambda(A\times B)=P((A\cup B)^2)-P(A^2)-P(B^2).$$
		\item If $C:=A\cap B\ne\emptyset$, then using $A_1=A\backslash C,\ B_1=B\backslash C$ we obtain
		$$A\times B=C^2\cup(A_1\times C)\cup(C\times B_1)\cup(A_1\times B_1).$$
  \end{enumerate}
	Then we have
	\begin{eqnarray*}
	2\lambda(A\times B)&=&2[\lambda(C^2)+\lambda(A_1\times C)+\lambda(C\times B_1)+\lambda(A_1\times B_1)]\\
	                   &=&P(A^2)+P(B^2)+P((A_1\cup B_1)^2)-2P(A_1^2)-2P(B_1^2)
	\end{eqnarray*}
	and thus $\lambda=\lambda_P$.
	\item For $A,B\in\mathcal A$ we have $\lambda (A\times B)^2\leq\lambda(A^2)\cdot\lambda(B^2)$.\\
	To prove this it is sufficient to apply the positivity condition to
	$$f=\chi(A)-\alpha\cdot\chi(B),\  \alpha\in\mathbb R$$

	and then to optimize the resulting inequality for $\alpha\in\mathbb R$.
\end{enumerate}
\vskip 3mm
The concept of a context $\mathcal K\subset\mathcal E$ is defined as a generalization from the finite $\Omega$ case. At first we define universal sets (universes) and then contexts.

\Definition
\begin{enumerate}[(i)]
	\item The event $U\in\mathcal E$ is a universe iff $\spt U=\Omega$
	\item Let $A=\bigcup_{\alpha\in I}A_\alpha^2,\ B=\bigcup_{\beta\in J}B_\beta^2\in\mathcal E$.\\
	We set $A\leq B$ iff $\forall\alpha\in I\ \exists\beta\in J$ such that $A_\alpha=B_\beta$.
	\item The set $\mathcal K\subset\mathcal E$ is a context if there exists a universe $U\in\mathcal E$ such that
	$$\mathcal K=\mathcal K_{U}:=\{E\in\mathcal E|E\leq U\}$$
\end{enumerate}
\Proposition Let $U=\bigcup_{\alpha\in I}A_\alpha^2$ is a universe and $\mathcal K=\mathcal K_{U}$ the corresponding context.
\begin{enumerate}[(i)]
	\item $A\in\mathcal K$ iff $\exists J\subset I$ such that
	$$A=\bigcup_{\alpha\in J}A_\alpha^2,\ A\in\mathcal E$$
	\item $A,B\in\mathcal K\Rightarrow A\pitchfork B$
	\item $\mathcal K$ is isomorphic to the standard Boole algebra $I$ by the maps $J\mapsto A$ defined in (i).
	\item Two different universes are incompatible
\end{enumerate}
\vskip 3mm
We see that $A_\alpha^2,\alpha\in I$ are, in fact, elementary events in $\mathcal K$. The canonical forms of $\mathcal K$ is given in the following definition.

\Definition Let $(\Omega^2,\mathcal E, P)$ be a quadratic probability space and $\mathcal K=\mathcal K_{U}\subset\mathcal E$ be a context.
\begin{enumerate}[(i)]
	\item The set of elementary events of $\mathcal K$ is given by
	$$\Omega_{\mathcal K}:=\{A_\alpha^2|\alpha\in I\}=\{A^2|A\in\mathcal A, A^2\in\mathcal K\}$$
	\item The algebra $\mathcal A_{\mathcal K}$ of events in $\mathcal K$ is defined by
	$$\{A_\alpha^2|\alpha\in J\}\in\mathcal A_{\mathcal K}\Leftrightarrow\bigcup_{\alpha\in J}A_\alpha^2\in\mathcal K,\ \forall J\subset I$$
	\item If $P(U)>0$ (i.e. $\mathcal K$ is $P$-regular), then we set
	$$\mathbb F_{\mathcal K}(E):=\frac{P(E)}{P(U)},\ E\in\mathcal K$$
\end{enumerate}
\vskip 3mm
\Proposition Let $\mathcal K=\mathcal K_{U}\subset\mathcal E$ be a context
\begin{enumerate}[(i)]
	\item $\mathcal A_{\mathcal K}$ is a $\sigma$-algebra
	\item $(\Omega_{\mathcal K},\mathcal A_{\mathcal K},\mathbb F_{\mathcal K})$ is the Kolmogorov probability space if $\mathcal K$ is $P$-regular
\end{enumerate}
\Proof (i) let
$$E_i=\{\bigcup_\alpha A_\alpha|\alpha\in I_i\}=\mathcal K,\ i=1,2,\dots$$
then we set $I:=\bigcup_{i=1}^\infty I_i$ and we obtain from $E_i\in(\overline{\mathcal A\times\mathcal A})$ that 
$$E:=\bigcup E_i=\bigcup_{\alpha\in I}A_\alpha^2\in(\overline{\mathcal A\times\mathcal A})\cap ST_{\Omega^2}.$$
Thus $E\in\mathcal E$.

(ii) The $\sigma$-aditivity of $\mathbb F_{\mathcal K}$ follows from the $\sigma$-additivity of $\lambda$.$\square$
\vskip 3mm

\Remark If $\Omega$ is finite we have the canonical algebra $\mathcal A_\Omega$ and the canonical counting measure $\nu_\Omega$ on $\Omega$, $\nu_{\Omega}(\mathcal A)=|A|$. Then we can define the probability distribution by
$${\bf p}=\frac{d\lambda}{d\nu_\Omega\times\nu_\Omega},\hbox{ i.e. }\lambda=(\nu_\Omega\times\nu_\Omega)\lefthalfcup{\bf p}.$$

In the case of general $\Omega$, there is no canonical measure $\nu_{\Omega}$. This gives the motivation of the following definition.

\Definition Let $\nu$ be a $\sigma$-finite measure on the algebra $\mathcal A$.
\begin{enumerate}[(i)]
	\item The quadratic  probability  space $(\Omega^2,\mathcal A, P)$ is $\nu$-regular iff $\lambda=\lambda_P$ is absolutely continuous with respect to $\nu\times\nu$ on $\overline{\mathcal A\times\mathcal A}$.
	\item $(\Omega^2,\mathcal E,P)$ is regular iff it is $\nu$-regular for some $\sigma$-finite measure $\nu$ on $\mathcal A$
	\item If $(\Omega^2,\mathcal E,P)$ is $\nu$-regular, then the Radon-Nikodym derivative
	$${\bf p}=\frac{d\lambda}{d\nu\times\nu}$$
	is called the probability distribution of $P$. (Of course, ${\bf p}$ depends on the choice of $\nu$.)\\
	Equivalently $\lambda$ is defined by ${\bf p}$
	$$\lambda=(\nu\times\nu)\lefthalfcup{\bf p}.$$
\end{enumerate}
\vskip 3mm
If we fix the $\sigma$-finite measure $\nu$ on $\mathcal A$, then it is possible to define the state space corresponding to $\nu$.

\Definition Let $\nu$ be a $\sigma$-finite meaure on $\mathcal A$.\\
The state space
$$\mathcal S(\Omega^2,\mathcal A,\nu)$$
is defined as a set of all functions
$${\bf p}:\Omega^2\rightarrow\mathbb R$$
(called probability distributions) which satisfy
\begin{enumerate}[(i)]
	\item ${\bf p}$ is symmetric: ${\bf p}(x,y)={\bf p}(y,x),\ \forall x,y\in\Omega$
	\item ${\bf p}$ is $\nu\times\nu$-integrable
	\item ${\bf p}$ is positive semi-definite, i.e.
	$$\int {\bf p}(x,y)f(x)f(y)d\nu\times\nu(x,y)\geq 0$$
	for each $f:\Omega\rightarrow\mathbb R$ which is $\nu$-integrable
	\item $\int_\Omega{\bf p}(x,x)d\nu(x)>0$
\end{enumerate}
\vskip 3mm
\Proposition Let ${\bf p}\in\mathcal S(\Omega^2,\mathcal A,\nu)$. If we set
\begin{enumerate}[]
	\item $\mathcal E=\overline{\mathcal A\times\mathcal A}\cap ST_{\Omega^2}$
	\item $P=(\nu\times\nu)\lefthalfcup{\bf p}$
\end{enumerate}
then $(\Omega^2,\mathcal E, P)$ is the quadratic probability space.

\Proof The proof is simple, only the last property $P(U)>0$ needs the more technical argument. This follows from the following theorem (part (iii)).
\vskip 3mm

\Theorem Let ${\bf p}\in\mathcal S(\Omega^2,\mathcal A,\nu)$ is a probability distribution

\begin{enumerate}[(i)]
	\item For $\nu$-a.e. $x\in\Omega$ and $\nu$-a.e. $y\in\Omega$ (a.e.=almost every) we have
	$${\bf p}(x,y)^2\leq{\bf p}(x,x)\cdot{\bf p}(y,y)$$
	\item If $f\in L^2(\Omega,\nu)$, $L^2$=real Hilbert space, then the integral
	$$\int{\bf p}(x,y)f(y)d\nu(y)$$
	exists and defines the operator
	$$\hat{\bf p}:L^2(\Omega,\nu)\rightarrow L^2(\Omega,\nu)$$
	and, moreover, we have
	$$|\hat{\bf p}(f)(x)|^2\leq{\bf p}(x,x)\cdot\tr{\bf p}\cdot ||f||_{L^2}^2,\ \forall\nu\hbox{-a.e. }x\in\Omega$$
	where
	$$\tr{\bf p}:=\int_\Omega{\bf p}(x,x)d\nu(x)$$
	and
	$$||\hat{\bf p}(f)||_{L^2}\leq\tr{\bf p}\cdot||f||_{L^2},$$
	$$||\hat{\bf p}||_{op}:=\sup\kern -8mm\lower 2mm\hbox{${}_{||f||_{L^2}=1}$}\ \ \ ||\hat{\bf p}(f)||\leq\tr{\bf p}.$$
	\item There exists $A\in\mathcal A$ such that
	$$\int_{A^2}{\bf p}\ d\lambda=\lambda_P(A^2)>0$$
\end{enumerate}

\Proof (i) From the theory of the derivation of measures it follows that  there exists sets $A_i^{z}\in\mathcal A,\ i=1,2,\dots,z\in\Omega$ such that  for $\nu$-a.e. $z\in\Omega$ and $\nu$-a.e. $w\in\Omega$ we have
$$\int{\bf p}(x,y)\varphi_i^z(x)\varphi_i^w(y)d\nu(z)d\nu(w)\ \ {\longrightarrow}\kern -6mm\lower 2mm\hbox{${}_{{i\rightarrow\infty}}$}\ \ \ {\bf p}(z,w)$$
where
$$\varphi_i^z(x):=\frac{\chi(A^z_i;x)}{\nu(A_i^z)}.$$
Using the positivity condition with
$$f=\varphi_i^z-\alpha\cdot\varphi_i^w,\ \ \alpha\in\mathbb R,$$
we shall find the optimal $\alpha\in\mathbb R$

(ii) We denote $f'(x)=\int{\bf p}(x,y)f(y)d\nu(y)$ and we have using (i)\
$$|f'(x)|^2\leq\int|{\bf p}(x,y)|^2d\nu(y)\cdot\int|f(y)|^2d\nu(y)$$
$$\leq{\bf p}(x,x)\int{\bf p}(y,y)d\nu(y)\cdot\int|f(y)|^2d\nu(y).$$
Then
$$\int|f'(x)|^2d\nu(x)\leq\left(\int{\bf p}(x,x)d\nu(x)\right)^2\cdot\int|f(y)|^2d\nu(y).$$
(iii) We have
$$\int{\bf p}(x,x)d\nu>0,\ {\bf p}(x,x)\geq 0,\ \forall\nu\hbox{-a.e. }x\in\Omega$$
From [1,3.1.2 Theorem 4] it follows that  the operator $\hat{\bf p}$ has spectral decomposition
$$p(x,y)=\sum_i \lambda_i\varphi_i(x)\varphi_i^*(y),\ \lambda_i\geq 0,\ ||\varphi_i||_{L^2}=1$$
with $\tr{\bf p}=\sum\lambda_i>0$. Let us assume that $\lambda_s>0$ since $\nu$ is $\sigma$-finite there exists $A\subset\Omega$ such that $\nu(A)<\infty$, $\int_A|\varphi_{s}|^2d\nu>0$.
Let us denote $\psi:=\varphi_{s}\cdot\chi(A)$. Then $\psi\in L^1(\Omega,\nu)$ and $\int_A|\psi|^2d\nu>0$.

Let us assume that $\int_B\psi d\nu=0,\ \forall B\subset A, B\in\mathcal A$. 

This means that $\nu\lefthalfcup\psi=0$ and then $\psi=0\ \nu$-a.e. but this contradicts to $\int_B|\psi|^2>0$.

We have obtained that there exists $B\subset A, B\in\mathcal A$ such that $\int_B\psi d\nu\ne 0$.

Then we have
$$\int_{B^2}=\varphi_{s}(x)\varphi_{s}(y)^*d\nu(x) d\nu(y)=|\int_B\varphi_{s}d\nu|^2>0.$$
For each $i$ we have
$$\int_{B^2}\varphi_i(x)\varphi_i(y)^*d\nu(x)d\nu(y)=|\int_B\varphi_id\nu|^2\geq 0$$
so that
$$\lambda_P(B^2)=\int_{B^2}{\bf p}(x,y)d\nu(x)d\nu(y)\geq|\int_B\varphi_{s}d\nu|^2>0.\hbox to 5mm{}\square$$
\vskip 3mm
\Remark There are equivalent formulations:
\begin{enumerate}[(i)]
	\item $\exists$ a universe $U, P(U)>0$
	\item $\exists E\in\mathcal E, P(E)>0$
	\item $\exists A\subset\Omega,\ A\in\mathcal A$ such that $P(A^2)>0$.
\end{enumerate}
\vskip 3mm
The concept of the observation of an individual system is classical: the observation shows which elementary event from $e_1,\dots, e_n$ has happened.

But we must take into account the basic fact, that each observation is well defined only if the context of this observation is specified.
\vskip 3mm

Let us assume that we are observing the system in the $P$-regular context $\mathcal K$ defined by its universe
$$U_{\mathcal K}=\bigcup_{\alpha\in I}A_\alpha^2,\ \bigcup A_\alpha=\Omega.$$
Elementary events in $\mathcal K$ are atoms $A_\alpha^2,\ \alpha\in I$. By observing the system in the context $\mathcal K$, we find which atomic event $A_\alpha^2,\ \alpha\in I$ has happened. (Regularity of $\mathcal K$ implies that  $P(A_\alpha^2)>0$ for some $\alpha\in I$, i.e. something will happen.)
\vskip 3mm

For example, if $\mathcal K$ is the classical context $\mathcal K^{cl}(\Omega\hbox{ finite})$, then we have
$$U_{\mathcal K^{cl}}=\bigcup_{i=1}^n\{e_i\}^2=\diag\Omega^2,\ \bigcup\{e_i\}=\Omega.$$
Observing the system in $\mathcal K^{cl}$ we find which elementary event $e_i$ has happened.
\vskip 3mm
A random variable $X$ is a quantity which value depends on the case - which event has happened. Thus $X$ is well defined only if the context is given.

\Definition Let $\mathcal K$ be a $P$-regular context and $U_{\mathcal K}=\bigcup_\alpha A_\alpha^2$ its universe.
\begin{enumerate}[(i)]
	\item the map
	$$X:U_{\mathcal K}\rightarrow \mathbb R$$ is a $\mathcal K$-random variable (i.e. $X\in R V_{\mathcal K}$) iff $X$ is constant on each atom $A_\alpha^2,\ \alpha\in I$ from $\mathcal K$
	\item $X$ is conventionally extended to $\Omega^2$ by
	$$X=0\hbox{ on }\Omega^2\backslash U_{\mathcal K}.$$
\end{enumerate}
\vskip 3mm
\Remark Clearly, $X$ is in fact the map
$$X:\{A_\alpha^2|\alpha\in I\}\rightarrow\mathbb R$$
$$X(A_\alpha^2)=X(z),\ \forall z\in A_\alpha^2.$$
Thus $X$ can be considered as a standard random variable on the classical probability space
$$(\Omega_{\mathcal K}, \mathcal A_{\mathcal K},\mathbb F_{\mathcal K})$$
defined above.
\vskip 2mm
\Proposition Let $X\in RV_{\mathcal K}$, $\mathcal K$ be a $P$-regular context. Then
\begin{enumerate}[(i)]
	\item $X$ is symmetric on $\Omega^2$
	\item when the experiment is repeated in $(\Omega^2,\mathcal E, P)$ and in the $P$-regular context $\mathcal K,\ U_{\mathcal K}=\bigcup_\alpha A_\alpha^2$ then the mean value of $X$ is given by
	$$\langle X\rangle_P:=\frac{1}{P(U_{\mathcal K})}\int_{\Omega^2} Xd\lambda_P$$
	where $\lambda_P$ is the signed measure on $\overline{\mathcal A\times\mathcal A}$ associated to $P$ and where it is assumed that $X$ is $\lambda_P$-integrable.
\end{enumerate}

\Proof (i) If $X(x,y)\ne 0,\ x\ne y$ then $(x,y)\in U_{\mathcal K}$. Then there exists $\beta$ such that $(x,y)\in A_\beta^2$ so that $(y,x)\in A_\beta^2$ and $X(x,y)=X(y,x)$ since $X$ is constant on $A_\beta^2$

(ii) the experiment is repeated in the classical probability model
$$(\Omega_{\mathcal K},\mathcal A_{\mathcal K},\mathbb F_{\mathcal K})$$
corresponding to the context $\mathcal K$ and defined above. We shall assume that $X\geq 0$ on $U_{\mathcal K}$ and on $\Omega_{\mathcal K}$. In the classical probability model we have (the Law of Large Numbers)
$$\langle X\rangle_{\mathbb F_{\mathcal K}}=\int_{\Omega_{\mathcal K}}Xd\mathbb F_{\mathcal K}=\frac{1}{P(U_{\mathcal K})}\int Xd\tilde{P}$$
where
$$\tilde{P}(B):=P(\bigcup_\alpha\{A_\alpha^2|A_\alpha\in B\}),\ \forall B\in\mathcal A_{\mathcal K}.$$
Since $X$ is constant on each $A_\alpha^2$ we have
$$\langle X\rangle_{\mathbb F_{\mathcal K}}=\frac{1}{P(U_{\mathcal K})}\int_{U_{\mathcal K}}Xd\lambda_P=\frac{1}{P(U_{\mathcal K})}\int_{\Omega^2}Xd\lambda_P$$
Then we set $X=X^+-X^-,\ X^+=\max(X,0)$ $\square$.
\vskip 3mm

If $P$ is $\nu$-regular, i.e. ${\bf p}=d\lambda_P/d\nu\times\nu\in\mathcal S(\Omega^2,\mathcal A,\nu)$ then
$$\langle X\rangle_P=\int_{\Omega^2}X\cdot{\bf p }d\nu\times\nu$$
\vskip 3mm

Thus each $X$ defines the convex functional
$$X:\mathcal S(\Omega^2,\mathcal A,\nu)\rightarrow\mathbb R\hbox{ by}$$
$${\bf p}\mapsto\langle X\rangle_{P_{\bf p}}\hbox{ where }{P_{\bf p}}:=\nu\times\nu\lefthalfcup{\bf p}.$$
I.e. each random variable can be considered as "linear" functional on the state space.
\vfill\eject

\bigskip As a Conclusion, the main points of our approach are the following :

\begin{enumerate}
\item Events are modelled as partitions, not as subsets. This is the novel
feature which make EPT completely different from CPT - CPT\ is obtained when
EPT is restricted to a context (in this sence CPT is a part of EPT).
\item Events have the quadratic structure represented by (+), while the
probability measure is additive.
\item The relation (+) follows from the structure of partitions and need not be
postulated as in QMT.
\item This is in contrast to QMT where events are subsets, but the quantum
measure has the quadratic structure.
\end{enumerate}
\bigskip

Modelling events as partitions introduces new concepts: incompatibility, contexts, quadratic
probability spaces. The resulting state space resembles the state space in the
so-called "real" quantum mechanics - the real density matrices. To each
experiment there is associated a context of all events observable in this
experiment.(There are always events not observable in a given experiment.) The
context represents (and realizes) the which-way information and this is
the way how the which-way information can enter into physics. In all paper we
consider events, partitions and ST-sets as different objects.

\bigskip

In the continuation it will be shown that EPT have many features similar to QM
and that QM\ can be represented in EPT as a standard Markov process. In this
way the Einstein%
\'{}%
s vision of QM\ as a stochastic theory will be realized.

\bigskip

\vfill\eject{\Large{\textbf{References.}}}
\begin{enumerate}[{[1]}]
	\item Teiko Heinosaari, Mario Ziman: \textit{Guide to mathematical concepts of quantum theory}, Acta Physica Slovaca, vol. 58, No. 4, 478--674, August 2008.
	\item Ji\v r\'{\i} Sou\v cek: \textit{The complex probability theory as a basis of quantum theory}, in Proceedings of Wint. School Abstr. Anal., \v Spindlerův Ml\'yn 1980, Math. Inst. Czech. Acad. Sci., Praha, 1980, pp. 151--154.
	\item Jiří Souček: arXiv: quant-ph/01071017 v1, Jul 2001
	\item R.~Sorkin: arXiv gr-qc/9401003, arXiv:1004.1226 and references therein
	\item S.~Gudder: arXiv:0909.223, arXiv:1005.2242 and references therein
\end{enumerate}
\end{document}